\documentclass[prb,twocolumn,preprintnumbers,amsmath,amssymb,superscriptaddress]{revtex4}
\usepackage{epsf}
\usepackage{graphicx}
\usepackage{bm}
\usepackage{amsmath}
\usepackage{color}
\usepackage{notes2bib}
\usepackage{epstopdf}
\usepackage{textcomp}

\bibnotesetup{
note-name = , use-sort-key = false }
\begin{document}


\title{General theory of topological Hall effect in systems with chiral spin textures}

\author{K.~S.~Denisov}
\email{denisokonstantin@gmail.com} \affiliation{Ioffe Institute, 194021 St.Petersburg, Russia}
\affiliation{Lappeenranta University of Technology, FI-53851 Lappeenranta, Finland}
\author{I.~V.~Rozhansky}
\affiliation{Ioffe Institute, 194021 St.Petersburg, Russia}
\affiliation{Lappeenranta University of Technology, FI-53851 Lappeenranta, Finland}
\author{N.~S.~Averkiev}
\affiliation{Ioffe Institute, 194021 St.Petersburg, Russia}
\affiliation{Lappeenranta University of Technology, FI-53851 Lappeenranta, Finland}
\author{E.~L\"ahderanta}
\affiliation{Lappeenranta University of Technology, FI-53851 Lappeenranta, Finland}

\begin{abstract}
We present a consistent theory of the topological Hall effect (THE) 
in 2D magnetic systems with disordered array of chiral spin textures, such as magnetic skyrmions. 
We focus on the scattering regime when the mean-free path of itinerant electrons exceeds the spin texture size,
and THE arises from the asymmetric carrier scattering on individual chiral spin textures. 
We calculate the resistivity tensor on the basis of the Boltzmann kinetic equation 
taking into account the
asymmetric scattering on skyrmions   
via the collision integral. 
Our theory describes both 
the adiabatic regime, when THE arises from 
a spin Hall effect 
and 
the non-adiabatic scattering when THE is 
due to purely charge transverse currents. 
We analyze the dependence of THE resistivity on a chiral spin texture structure, as well as on 
material parameters.
We discuss the crossover between spin and charge regimes of THE driven by the increase of skyrmion size, 
the features of THE due to the variation of the Fermi energy, 
and the 
exchange interaction strength; 
we comment on the sign and magnitude of THE. 

\end{abstract}


\date{\today}

\maketitle

\section{Introduction}

Among rich variety of transport phenomena in magnetic materials 
the special attention is now focused on the topological Hall effect (THE). 
THE is the appearance of an additional transverse voltage  
due to itinerant carrier exchange interaction with chiral spin textures, such as magnetic skyrmions~\cite{NagaosaNature,fert2013skyrmions,fert2017magnetic,wiesendanger2016nanoscale}. 
THE has been extensively studied experimentally~\cite{ahadi2017evidence,room_T,soumyanarayanan2017tunable,raju2017evolution,B20_FeCoGe,MnSiAPhase}
and has proved itself as an indicator of 
a non-zero chirality of the sample magnetization.
The observation of THE has been reported for various systems exhibiting different  chiral ordering of spins: skyrmion crystals~\cite{MnSiAPhase,B20_FeCoGe,THE_Li,Chapman_PRB}, antiferromagnets (AFM)~\cite{surgers2014large,ueland2012controllable}, spin glasses~\cite{SpinGlass1,SpinGlass2} and disordered arrays of magnetic skyrmions~\cite{EuO,ahadi2017evidence,maccariello2018electrical,room_T,soumyanarayanan2017tunable,AronzonRozh,THE_TI,DiscretHall}.

Naturally, an appropriate microscopic theory of THE has to take into account the particular type of chiral spin ordering. 
In the case of a regular non-collinear spin structure with periodic or quasi-periodic spin arrangement, such as AFM latices~\cite{surgers2014large} or skyrmion crystals~\cite{B20_FeCoGe,Muhl_MnSi_Science,Munzer_PRB}, 
THE can be described in terms of an effective mean magnetic 
field~\cite{BrunoDugaev,B20_FeCoGe,binz2008chirality}. 
The meanfield approach is usually 
justified within the adiabatic approximation typical 
for systems with strong exchange coupling between the magnetization and the  electrons carrying the current~\cite{Ye1999,Lyana-Geller1,Arab_Papa,Tatara_2007,Diffusive_THE,Taguchi_Science}. 
Deviations from the condition of adiabaticity in THE due to  
long-ranged spin textures has been recently discussed in Ref.~\cite{nakazawa2018topological,ishizuka2018spin}. 




Another type of the chiral magnetization profiles studied experimentally  
is a disordered array of localized  small spin textures spatially separated from each other~\cite{soumyanarayanan2017tunable,maccariello2018electrical,denisov2018hall}. 
In this case a carrier moves freely most of the time, and the presence of localized magnetization vortices affects its trajectory by occasional scattering. 
The mean field approach is not adequate in this case as there is no regular long-range chiral spin structure which can be described by a homogeneous effective magnetic field.  
Instead, THE 
is driven by an asymmetric scattering of the carriers on individual spin textures   being sensitive to their particular magnetization profile. 
The important feature of the individual scattering regime is that the properties of  THE strongly depend on whether the carrier spin-flip processes are activated or not corresponding to so-called weak coupling regime and adiabatic regime, respectively~\cite{SciRep_skyrmion}. 
The transverse electric response arises 
from the spin Hall effect in the adiabatic regime and 
from the charge Hall effect in the weak coupling regime~\cite{prl_skyrmion,Tatara,Nakazawa2014,Taguchi_Science,nakazawa2018topological,Onoda_SkyrmNumber}.  
Thus, the complete theory of THE for the irregular dilute chiral systems
requires an accurate treatment of carrier scattering on a single chiral spin texture. 

In this paper we develop the 
theory of the topological Hall effect in the diffusive regime
for the dilute systems of localized magnetic textures. 
The  approach is based on the consideration of  asymmetric carrier scattering on individual spin textures.
Our theory is applicable to 
disordered arrays of chiral spin textures 
with both electron spin subbands populated  
when THE can be generated 
both by charge and spin transverse currents. 
The paper is organized as follows: 
in section \ref{s_2} the kinetic theory of THE is described accounting for the carrier scattering on host impurities and non-collinear spin textures, in section \ref{s_3} the properties of the exhange asymmetric scattering are discussed,  
section \ref{s_4} covers 
the dependence of THE on material and spin texture parameters, 
we also describe the crossover between charge and spin Hall regimes of THE driven by the suppression of spin-flip scattering; 
in section \ref{s_5} we summarize our results.

\section{Kinetic theory}
\label{s_2}
Let us consider two-dimensional degenerate electron gas (2DEG)
described by the Hamiltonian:
\begin{equation}
\label{eq_2DEG}
\begin{aligned}
& \mathcal{H} = \frac{p^2}{2 m}  -\alpha_{0} \boldsymbol{S}(\boldsymbol{r}) \cdot \boldsymbol{\sigma} +  \sum_i u(\boldsymbol{r}-\boldsymbol{r}_i)
\end{aligned}
\end{equation}
where the first term describes 
the electron free motion with an effective in-plane mass $m$, 
the second term represents the electrons exchange interaction with 
a magnetic texture described by a static spin field $\boldsymbol{S}(\boldsymbol{r})$, where 
$\alpha_0$ is an exchange coupling constant, $\boldsymbol{\sigma}$ is the vector of Pauli matrices, 
the last term describes scattering on host non-magnetic impurities located at $\boldsymbol{r}_i$. 
The topological Hall effect appears 
when $\boldsymbol{S}(\boldsymbol{r})$ has a non-collinear structure characterized by a non-zero spin chirality. 

We consider the case, when 
the spin field $\boldsymbol{S}(\boldsymbol{r})$ consists of 
two contributions:
\begin{equation}
    \label{eq_S}
    \boldsymbol{S}(\boldsymbol{r}) = S_0 \boldsymbol{e}_z + \sum_j \delta \boldsymbol{S}(\boldsymbol{r}-\boldsymbol{r}_j).
\end{equation}
The first term is a background homogeneous field directed perpendicular to the 2DEG plane leading to the Zeeman spin splitting $\Delta = \alpha_0 S_0$. 
We assume ferromagnetic exchange ($\Delta>0$) and that the Fermi energy exceeds the spin splitting so that both spin subbands are populated. 
The second contribution describes localized non-collinear spin textures $\delta \boldsymbol{S}$ of a few nanometer size located at $\boldsymbol{r}_j$ and causing an additional elastic scattering of the carriers. 
While magnetic skyrmions are the typical example of a localized non-collinear spin texture,
our consideration covers much wider class of 
chiral spin textures, not necessarily having a non-zero topological charge~\cite{SciRep_skyrmion}.
The feature of the chiral spin structures 
is that 
for a given incident electron flux 
there is a difference in scattering rates to the 
left and to the right, eventually leading to the Hall effect.

We consider the classic transport regime ($k_F \ell \gg 1$ where $k_F = \sqrt{2m E_F}/\hbar$ is Fermi wave-vector, $\ell$ is the mean free path) on the basis of the Boltzmann kinetic equation:
\begin{equation}
\label{eq_B}
\begin{aligned}
& e \boldsymbol{E} \cdot \frac{\partial f_s(\boldsymbol{p})}{\partial \boldsymbol{p}} = {\rm St}\left[ f_s(\boldsymbol{p}) \right], \\
& {\rm St}\left[ f_s(\boldsymbol{p}) \right]  = 
\sum_{\boldsymbol{p}',s'} \left( \mathcal{W}_{\boldsymbol{p p}'}^{ss'} f_{s'}(\boldsymbol{p'}) - \mathcal{W}_{\boldsymbol{p' p}}^{s's} f_{s}(\boldsymbol{p}) \right), 
\end{aligned}
\end{equation}
where $f_s(\boldsymbol{p})$ is the distribution function, $\boldsymbol{p}$ is 2D momentum and $s=\pm1/2$ is the carrier spin projection on the axis normal to the  motion plane, $\boldsymbol{E}$ is an in-plane electric field, $\mathcal{W}_{\boldsymbol{p p}'}^{ss'} $ is the elastic scattering rate from ($\boldsymbol{p}'$, $s'$) to ($\boldsymbol{p}$, $s$) state, and $e$ is the electron charge. 
We solve Eq.(\ref{eq_B}) in linear approximation with respect to $\boldsymbol{E}$. 

Expressing the scattering rate $\mathcal{W}_{\boldsymbol{p p}'}^{ss'} $ in the form of the Fermi Golden Rule we assume that it has   
 two contributions:
\begin{equation}
\label{eq_rates}
\begin{aligned}
& \mathcal{W}_{\boldsymbol{p p}'}^{s s'} = 
\frac{2\pi}{\hbar} \left( n_i |u_{\boldsymbol{p}\boldsymbol{p}'}|^2 \delta_{ss'} + n_{sk} |T_{\boldsymbol{p}\boldsymbol{p}'}^{ss'}|^2 \right) \delta\left(\varepsilon_p^s - \varepsilon_{p'}^{s'}\right),
\end{aligned}
\end{equation}
where the first term in brackets describes the electron spin-independent scattering on non-magnetic impurities, the second term is driven by the scattering on chiral spin textures; interference effects between the two types of scatterers are neglected. 
Here 
$n_i$, $n_{sk}$ are the sheet densities of impurities and localized magnetic textures, respectively, 
$u_{\boldsymbol{p}\boldsymbol{p}'}$ is Fourier transform of the non-magnetic impurity potential $u(\boldsymbol{r})$ from Eq.~(\ref{eq_2DEG}),  $T_{\boldsymbol{p}\boldsymbol{p}'}^{ss'}$ is the exact $T$-matrix of electron scattering on the spin texture, and 
the delta-function ensures energy conservation in the elastic scattering,  the energy spectrum is $\varepsilon_p^s = {p^2/2m - s \Delta}$.
Two contributions can be distinguished in the square modulus of the $T$-matrix:  
\begin{equation}
    \begin{aligned}
    & |T_{\boldsymbol{p}\boldsymbol{p}'}^{ss'}|^2 = \frac{1}{\nu^2} \left( \mathcal{G}_{ss'}(\theta) + \mathcal{J}_{ss'}(\theta) \right),
    \\
    & \mathcal{G}_{ss'}(\theta) = \mathcal{G}_{ss'}(-\theta) 
    \hspace{0.5cm}
    \mathcal{J}_{ss'}(\theta) =- \mathcal{J}_{ss'}(-\theta).
\end{aligned}
\label{eq_Important}
\end{equation}
$\mathcal{G}_{ss'}(\theta)$, $\mathcal{J}_{ss'}(\theta)$ are dimensionless symmetric and asymmetric scattering rates, respectively, $\theta$ is the scattering angle,
$\nu = m/2\pi \hbar^2$ is 2D density of states (per one spin). 
In the introduced notation we omit the dependence of $\mathcal{G}_{ss'},\mathcal{J}_{ss'}$ on the scattering energy.  

It is the asymmetric part $ \mathcal{J}_{ss'}(\theta)$ of an electron scattering on chiral spin textures that gives rise to the transversal current as the scattering rates to the left and to the right 
become unequal. The scattering asymmetry acts as an effective magnetic field, which sign can be 
either the same for both spin projections of an incident electron, hence leading to a charge Hall effect, or 
opposite for the opposite electron spin projections, leading to the spin Hall effect. 
The properties of $\mathcal{J}_{ss'}(\theta)$ are discussed in 
Ref.~\cite{SciRep_skyrmion} and summarized in Section~\ref{s_3}.

In order to solve the kinetic equation (\ref{eq_B}) 
we write $f_s(\boldsymbol{p}) = f_s^{0} + g_s(\boldsymbol{p})$, where $f_s^{0}$ and $g_s$ are the equilibrium and non-equilibrium parts of the distribution function, respectively. 
The external electric field $\boldsymbol{E}$ is directed along $x$-axis. 
The angular dependence of $g_s(\boldsymbol{p})$ can be expressed as a sum of
two terms, even and odd with respect to the angle:
\begin{equation}
\label{eq_f}
g_s(\boldsymbol{p}) = g_s^{+}(p) \cos{\varphi} + g_s^{-}(p) \sin{\varphi},
\end{equation}
where $\varphi$ is the momentum $\boldsymbol{p}$ polar angle counted from $x$ axis.
After integrating the collision integral (\ref{eq_B}) over $\boldsymbol{p}'$
we arrive at the following system of equations: 
\begin{equation}
\label{eq_sys}
eE \begin{pmatrix}
v_{\uparrow} \frac{\partial f_{\uparrow}^{0}}{\partial \varepsilon}\\
0\\
 v_{\downarrow} \frac{\partial f_{\downarrow}^{0}}{\partial \varepsilon}\\
0\\
\end{pmatrix}
=
\begin{pmatrix}
-{\tau_{\uparrow}^{-1}} & \Omega_{\uparrow \uparrow} & {\tau_{\uparrow \downarrow}^{-1}} & - \Omega_{\uparrow \downarrow}
\\
-\Omega_{\uparrow \uparrow} & -{\tau_{\uparrow}^{-1}} & \Omega_{\uparrow \downarrow} & {\tau_{\uparrow \downarrow}^{-1}}
\\
{\tau_{\downarrow \uparrow}^{-1}} &   - \Omega_{\downarrow \uparrow}  & -{\tau_{\downarrow}^{-1}} &  \Omega_{\downarrow \downarrow}
\\
\Omega_{\downarrow \uparrow}  & {\tau_{\downarrow \uparrow}^{-1}}& - \Omega_{\downarrow \downarrow} & -{\tau_{\downarrow}^{-1}}
\end{pmatrix}
\begin{pmatrix}
g_{\uparrow}^{+}\\
g_{\uparrow}^{-}\\
g_{\downarrow}^{+}\\
g_{\downarrow}^{-}\\
\end{pmatrix}.
\end{equation}

Here we introduced the following parameters:
\begin{equation}
\label{eq_rate}
\begin{aligned}
& \tau_s^{-1} = \tau_0^{-1} + \omega_{s};
\hspace{0.4cm}
\tau_0^{-1} = n_i \frac{2\pi}{\hbar}  \nu  \int\limits_0^{2\pi} |u_{\boldsymbol{p}\boldsymbol{p}'}|^2 (1-\cos{\theta)} \frac{d\theta}{2\pi};
\\
& \omega_s = n_{sk} \frac{2\pi}{\hbar} \int\limits_0^{2\pi} \left[ \left( 1 - \cos{\theta} \right) \mathcal{G}_{ss}(\theta) + \mathcal{G}_{\bar{s}s} \right] \frac{1}{\nu} \frac{d\theta}{2\pi};
\\
&\tau_{s\bar{s}}^{-1} = n_{sk} \frac{2\pi}{\hbar} \int\limits_0^{2\pi} \mathcal{G}_{s\bar{s}}(\theta) \cos{\theta} \frac{1}{\nu} \frac{d\theta}{2\pi};\\
& \Omega_{ss'} = \frac{e B_{ss'}}{mc};
\hspace{0.5cm}
B_{ss'} = (n_{sk} \phi_0) \int\limits_0^{2\pi} \mathcal{J}_{ss'}(\theta) \sin{\theta} d\theta,
\end{aligned}
\end{equation}
where $\tau_s$ is the total transport lifetime, $\tau_0$ is the transport lifetime for the scattering on non-magnetic impurities, 
$\omega_{s}^{-1}$ and $\tau_{s\bar{s}}$ are the transport lifetimes for the scattering on chiral textures (here $\bar{s}$ is the spin subband index opposite to $s$). 
The transverse Hall current due to the asymmetrical scattering is related to the parameter $\Omega_{ss'}$, which is analogous to the cyclotron frequency 
in the ordinary Hall effect; 
$B_{ss'}$ is the corresponding effective magnetic field, $\phi_0 = hc/|e|$ is the magnetic flux quantum, $c$ is the speed of light.  
 

The coefficients $g_s^{\mu}$ ($\mu=\pm$) linearly depend on the electric field: 
$g_{s}^{\mu} = {{\mathcal{A}}}_s^{\mu} E$, 
where ${\mathcal{A}}_{s}^{\mu}$ can be obtained from Eq.~(\ref{eq_sys}) by inverting the collision integral matrix. 
The longuitudonal $\sigma_{xx}^s$ and transverse $\sigma_{xy}^s$ conductivities within each spin subband $s$ are given by:
\begin{equation}
\label{eq_sig_g}
\begin{aligned}
& \sigma_{xx}^s = e \nu \int d\varepsilon \sqrt{\frac{\varepsilon}{2m}} \mathcal{A}_s^{+}(\varepsilon),
\\
& \sigma_{yx}^s = e \nu \int d\varepsilon \sqrt{\frac{\varepsilon}{2m}} \mathcal{A}_s^{-}(\varepsilon), \\
&  \sigma_{xx} = \sigma_{xx}^{\uparrow} + \sigma_{xx}^{\downarrow}, 
\hspace{0.5cm}
\sigma_{yx} = \sigma_{yx}^{\uparrow} + \sigma_{yx}^{\downarrow}, 
\hspace{0.5cm}
\hat{\rho} = \hat{\sigma}^{-1};
\end{aligned}
\end{equation}
where the integration goes over the energy $\varepsilon$, 
$\hat{\rho}$ is the tensor of resistivity. 
The topological Hall effect is determined by ${\mathcal{A}}_{s}^{-}$. 
Let us stress out that both spin-conserving and spin-flip scattering channels contain asymmetric parts $\Omega_{ss'}$ and thus contribute to ${\mathcal{A}}_{s}^{-}$, and THE. 
Moreover, as there exist different regimes of THE~\cite{SciRep_skyrmion}, it might be necessary to calculate the exact  
scattering rates $\mathcal{J}_{ss'}$. 


\section{Asymmetric electron scattering on a chiral spin texture}
\label{s_3}

In this section we consider the features of asymmetric electron scattering on a single non-collinear magnetic texture. 
We express the scattering potential in the form:
\begin{equation}
\label{eq_scattering}
\begin{aligned}
& V(\boldsymbol{r}) = - \alpha_{0} \delta\boldsymbol{S}(\boldsymbol{r}) \cdot \boldsymbol{\sigma}. \\
\end{aligned}
\end{equation}
Outside the localized spin texture of a characteristic diameter $a$ the magnetization is unperturbed so that  $\delta \boldsymbol{S}(r>a/2) \to 0$ and the scattering potential vanishes. 
The topological Hall effect appears due to the asymmetry in the electron scattering when the potential (\ref{eq_scattering}) can be characterized by a non-zero chirality. 
The details of the asymmetric scattering depend on the particular distribution of spins in the texture and its size as well as on the exchange interaction strength and the incident electron wavevector~\cite{SciRep_skyrmion}. 
%

\subsection{Chiral spin textures}

\begin{figure}
	\centering	
	\includegraphics[width=0.5\textwidth]{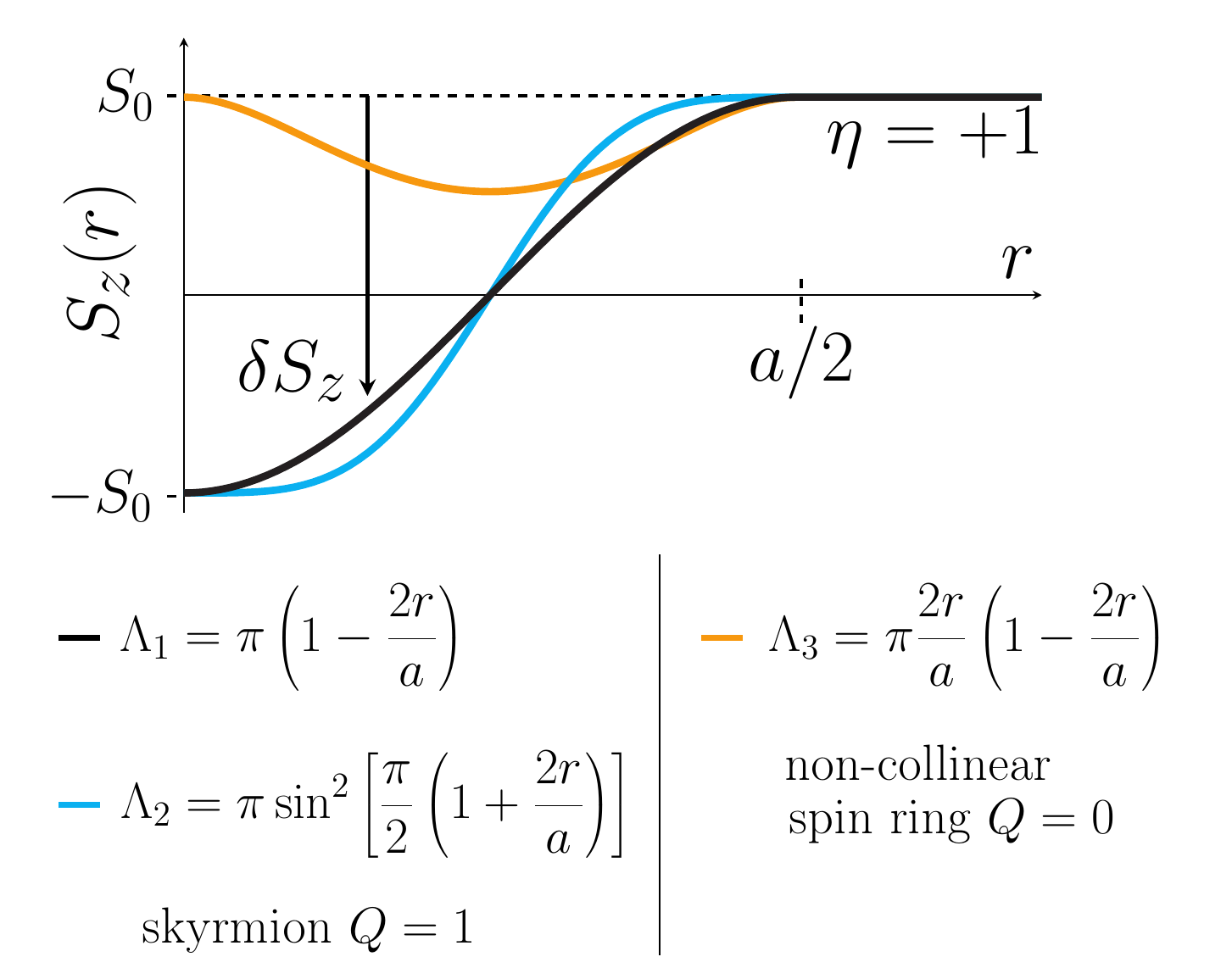}
	\caption{The typical profiles $S_z(r) = S_0 \cos{\Lambda(r)}$ of non-collinear spin textures. Note that for $\eta=+1$, the $\delta S_z(r) <0$ is negative.}	
	\label{f3}
\end{figure}

To describe a non-collinear spin texture in 2D  
the following parametrization is commonly used: 
\begin{equation}
\label{eq_deltaS}
\delta \boldsymbol{S}(\boldsymbol{r}) = 
\begin{pmatrix}
\delta S_{\parallel}(r) \cos{\left( \varkappa \phi + \gamma \right)}\\
\delta S_{\parallel}(r) \sin{\left( \varkappa \phi + \gamma \right)}\\
\delta S_{z}(r) 
\end{pmatrix}
\end{equation}
where $\boldsymbol{r} = (r,\phi) $ is the polar radius-vector, $r=0$ corresponds to the center of the texture.
The functions $\delta S_{\parallel}, \delta S_z \neq 0$ depend on the distance from the center $r$. 
The vorticity $\varkappa$ describes the in-plane spin rotation with an initial phase $\gamma$.
In what follows we consider that $\delta S_z$ is counted from the background magnetization $S_0$, whose sign we denote as $\eta = {\rm sgn}(S_0)$.
In the last section we will also consider the case when $S_0$ and $\delta S_z$ are independent.

Fig.~\ref{f3} shows  
the profiles $S_z(r) = S_0 \cos{\Lambda(r)}$ for three examples of chiral spin textures with $\eta = +1$ (we assume that $\delta S_{\parallel}^2 = S_0^2 - S_z^2 $). 
Two of them  describe a magnetic skyrmion ($\Lambda_1(r) = \pi(1-2r/a)$,  $\Lambda_2(r) = \pi \sin^2{\left[(\pi/2)(1+2r/a)\right]}$). 
The skyrmion has opposite sign of spins in its center with respect to the background magnetization, which leads to the appearance of a nonzero topological charge called winding number $Q$. The non-zero $Q$ is particularly important for the thermal stability of skyrmions in ferromagnetic thin films~\cite{Uzdin-1,moreau2016additive,soumyanarayanan2016emergent,bogdanov1994thermodynamically,bogdanov1999stability,belavin1975metastable,brey2017magnetic,bessarab2018lifetime}. 
The third magnetization profile Fig.~\ref{f3} ($\Lambda_3(r) = (2r/a)\pi(1-2r/a)$)) 
corresponds to  a non-collinear spin ring with the orientation of spins in the center parallel to $S_0$. Non-collinear rings have zero  winding number, but they exhibit a similar topological Hall effect~\cite{SciRep_skyrmion}. 
Such spin textures can appear in a material with with spin-orbit interaction functionalized by magnetic impurities,
in a vicinity of a defect or impurity~\cite{denisov2018hall,DenisovPolaron,Berk,Merkulov_Polaron}. 
Let us notice that for the positive background spin orientation ($\eta = +1$)  $\delta S_z$ is negative.

Substituting (\ref{eq_deltaS}) into (\ref{eq_scattering}) 
we get for the scattering potential:
\begin{equation}
\label{eq_Vsk}
V_{}(\boldsymbol{r}) = -\alpha_{0}
\begin{pmatrix}
 \delta S_z(r) & e^{- i \varkappa \phi - i \gamma} \delta S_{\parallel}(r)
\\
e^{ i \varkappa \phi + i \gamma} \delta S_{\parallel}(r)  & - \delta S_z(r)
\end{pmatrix}.
\end{equation}
The potential $V(\boldsymbol{r})$ is a $2 \times 2$ matrix, which depends on a polar angle $\phi$ via the off-diagonal components. 
When both functions $S_z, S_{\parallel}$ are non-zero, the angular dependence of the potential 
leads to the appearance of the asymmetric part 
in electron scattering rates 
$\mathcal{J}_{ss'}(\theta) = -\mathcal{J}_{ss'}(-\theta)$, 
where $\theta$ is the scattering angle. The sign of $\mathcal{J}_{ss'}$ depends on $\varkappa$, while the constant phase $\gamma$ plays no role in the scattering.
The role of $\eta$ is more complicated; 
we further explicitly specify the dependence of $\mathcal{J}_{ss'}$ on $\eta$.

\subsection{Asymmetric scattering features}


We consider the case when Fermi energy exceeds the background exchange splitting $E_F > \Delta/2$ 
so that both spin subbands are populated with electrons ($\Delta = \alpha_0 S_0$). 
The symmetry upon the time inversion 
allows us to present 
the asymmetric scattering rates $\mathcal{J}_{ss'}(\theta,\eta)$ introduced in Eq.~(\ref{eq_Important}) in the form
(see the details in Appendix~\ref{Ap1}):
\begin{equation}
    \label{eq_j}
    \begin{aligned}
& \mathcal{J}_{\uparrow \uparrow}(\theta,\eta) =  \eta  \varGamma_1(\theta) + {\Pi}(\theta),\\
& \mathcal{J}_{\downarrow \downarrow }(\theta,\eta) =  \eta \varGamma_1(\theta) - \Pi(\theta),\\
& \mathcal{J}_{\uparrow \downarrow }(\theta,\eta) = \mathcal{J}_{\downarrow \uparrow}(\theta,\eta)  =  \eta \varGamma_2(\theta),
    \end{aligned}
\end{equation}
where $\varGamma_{1,2}(\theta)$ and $\Pi(\theta)$ have no dependence on $\eta$. 
This representation is convenient for treating the topological charge and spin Hall effects independently. 
Indeed, the terms $\eta \varGamma_{1,2}$ describe the asymmetric scattering in the same transverse direction 
determined by the texture orientation $\eta$ and independent of an initial carrier spin state.
These terms, therefore, lead to the charge Hall effect. 
On the contrary, the term $\Pi$ describes the scattering of spin up and spin down electron in the opposite transverse directions independent of $\eta$. 
This process leads to spin Hall effect, it is absent for spin-flip channels. 
Both $\varGamma_{1,2}$ and $\Pi$ change their sign upon $\varkappa \to -\varkappa$.


Which of the two contributions to the topological Hall effect (charge or spin) dominate
strongly depends on whether the spin-flip processes are activated or not.
Away from the threshold $E_F \gg \Delta/2$ the rate of the spin-flip scattering is controlled by the adiabatic parameter 
$\lambda_a = (\alpha_{0} S_0/\hbar) \tau_{\rm a}$, where $\tau_{\rm a} = a/v_F$ is an electron time of flight  through the texture of diameter $a$ with Fermi velocity $v_F = \sqrt{2 E_F/m}$. 


In the case of $\lambda_a \le 1$ the spin-flip processes are effective, 
the asymmetric scattering arises from the interference between double spin-flip and single spin-conserving scattering events (so-called spin-chirality driven mechanism~\cite{prl_skyrmion,Tatara,nakazawa2018topological,Nakazawa2014,Onoda_SkyrmNumber}). This process is sensitive to the spin chirality $\chi$ 
defined for any three spins $\delta \boldsymbol{S}_1,\delta \boldsymbol{S}_2,\delta \boldsymbol{S}_3$ forming the spin texture as
$\chi=\text{sgn}\left(\delta \boldsymbol{S}_1 \cdot \left[ \delta \boldsymbol{S}_2 \times \delta \boldsymbol{S}_3 \right]\right)$.
The non-zero chirality of the spin texture in the weak coupling regime leads to the charge Hall effect.
The spin chirality based contribution is described by $\varGamma_{1,2}$. At $\lambda_a \le 1$ these terms dominate  $\varGamma_{1,2} \gg \Pi$, with spin-flip scattering prevailing $\varGamma_2 = 2 \varGamma_1$.


In the opposite case of large adiabatic parameter $\lambda_a \gg 1$ the spin flip processes are suppressed in accordance with the adiabatic theorem. 
In this regime the scattering asymmetry is due to the Berry phase acquired by the wave-function of electron moving through a non-collinear spin field. 
The hallmark of this mechanism is that the sign of the effective magnetic field 
associated with the Berry phase appears to be opposite for spin up and spin down electrons, thus leading to the spin Hall effect~\cite{BrunoDugaev,Taguchi_Science,Arab_Papa,Diffusive_THE}.
This adiabatic contribution to the Hall response is, therefore, described by $\Pi$. At $\lambda_a \gg 1$ the spin Hall effect dominates $\Pi \gg \varGamma_{1,2}$, and the charge Hall effect appears only due to nonzero carrier spin polarization $P_s$.


The interplay between charge and spin topological Hall effects leads to a few nontrivial
features
discussed in the following section.

\section{Topological Hall effect}

\label{s_4}

In this section we discuss the topological contribution to the Hall resistivity $\rho_{yx}^T$ 
in the diffusive regime for different systems.

\subsection{Dilute array of chiral spin textures}
\label{sb_crossover}

\begin{figure}
	\centering	
	\includegraphics[width=0.5\textwidth]{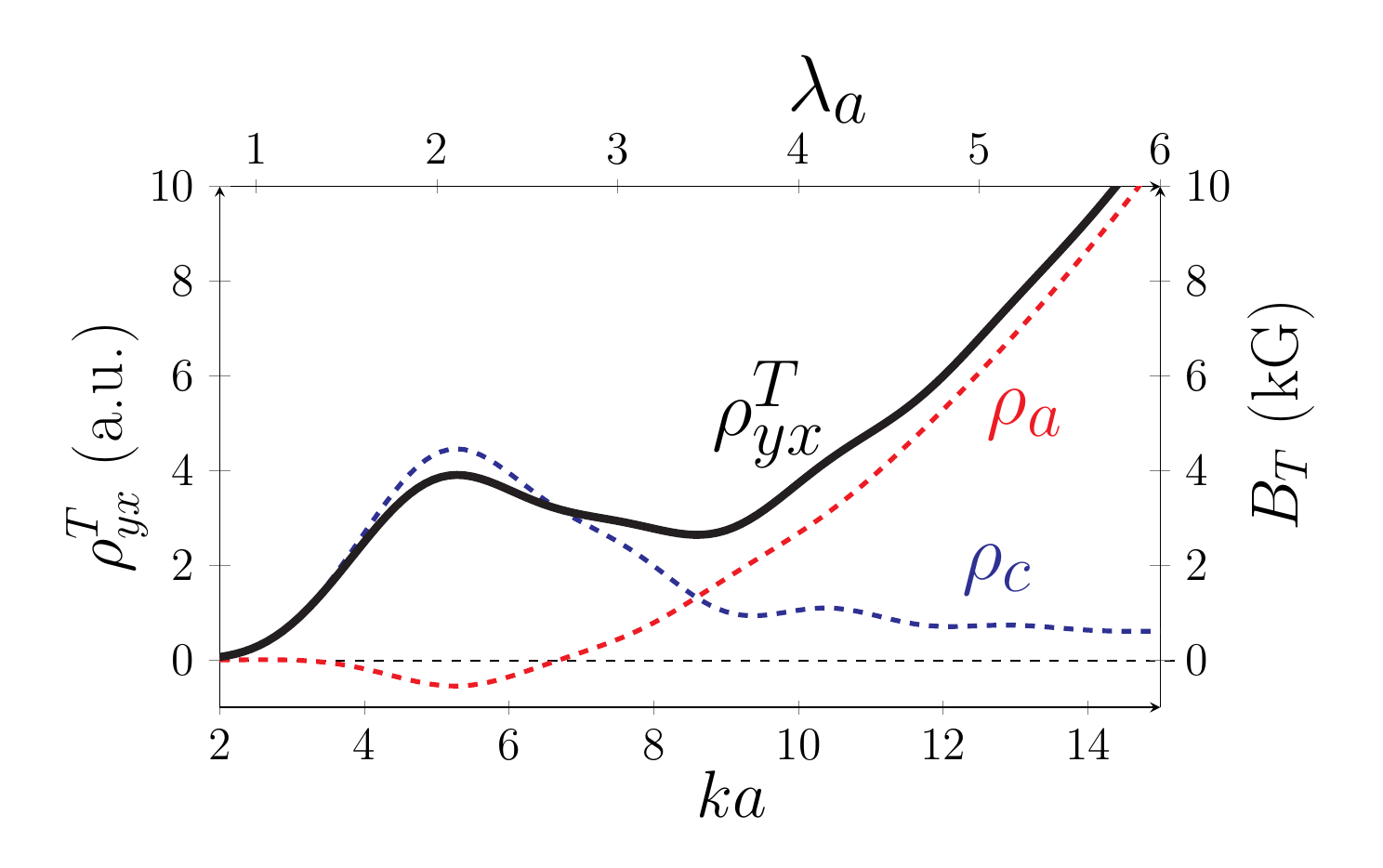}
	\caption{The dependence of $\rho_{yx}^T$ on magnetic skyrmion diameter $ka$ for $\Lambda_1$ profile, and the crossover between charge and spin topological Hall effect. The parameters $P_s =0.4$, $n_{sk} = 2\times 10^{11}$ cm$^{-2}$.}	
	\label{f1}
\end{figure}

Let us consider a two-dimensional film containing spatially localized chiral spin textures such as magnetic skyrmions or non-collinear magnetic rings (see Fig.~\ref{f3}). 
We assume that all the textures have the same vorticity $\varkappa$, and the orientation $\eta = {\rm sgn}(S_0) = +1$ is fixed, being determined by the background magnetization $S_0$. 
We consider the dilute regime, when the scattering rate on spin textures is much smaller than that on non-magnetic impurities $\omega_s \tau_0 \ll 1$, $\Omega_{ss'} \tau_0 \ll 1$, so the transport life-time is given by $\tau_s = \tau_0$. 
Solving the system (\ref{eq_sys}) for $\mathcal{A}_s^{-}$ in the lowest order in  $\left(\Omega_{ss'} \tau_0 \right)$ we express 
the topological Hall resistivity $\rho_{yx}^T$ as a sum of two contributions:
\begin{equation}
\label{eq_rhoT}
\begin{aligned}
& \rho_{yx}^T = \rho_{c} + \rho_{a}, \\
& \rho_{c} = \frac{1}{nec} \left(\phi_0 n_{sk}\right) \int\limits_0^{2\pi} \left( \varGamma_1 + \varGamma_2 \right) \sin{\theta} d\theta; 
\\
&\rho_{a} =  P_s \frac{1}{nec} \left(\phi_0 n_{sk}\right) \int\limits_0^{2\pi} \Pi \sin{\theta} d\theta. 
\end{aligned}
\end{equation}

The term $\rho_c$ describes the charge transverse current (charge Hall effect) generated due to carrier asymmetric scattering due to spin-independent terms $\varGamma_{1,2}$ (see Eq.~\ref{eq_j}).
The term $\rho_a$ describes the transverse spin current (spin Hall effect) driven by
the spin-dependent contribution to the asymmetric scattering $\Pi$ (Eq.~\ref{eq_j}).
The spin current does not lead to a charge separation unless there is unequal number of 
spin up and spin down carriers in the system. Therefore, this contribution to the Hall resistivity is proportional to the carrier spin polarization $P_s = (n_{\uparrow} - n_{\downarrow})/ (n_{\uparrow} + n_{\downarrow}) = \Delta/2E_F$. 
In Eq.~\ref{eq_rhoT} the notation $n$ stands for 2DEG sheet density. 

The relative importance of the two contributions  $\rho_a$ and $\rho_c$ in the appearance of the transverse charge current depends on the 
texture diameter $a$ or the Fermi level $E_F$ as discussed in the following sections.




\subsubsection{Crossover between charge and spin Hall effect}


Let us trace the dependence of $\rho_{yx}^T$  (\ref{eq_rhoT}) on the spin texture diameter $a$. 
We assume that the Fermi energy $E_F$ substantially exceeds the exchange spin splitting so that both 
spin subbands are populated and the spin polarization of the carriers is far below 100\%: 
$P_s = \Delta/2E_F \ll 1$.
In this case 
the rate of spin-flip processes are fully controlled by $\lambda_a$.
The adiabatic parameter can be expressed  as $\lambda_a = P_s (ka)$, 
where $k= \sqrt{2 E_Fm/\hbar^2}$.
When spin-flip scattering is activated 
the topological Hall effect is dominated by the transverse charge currents ($\rho_c \gg \rho_a$ at $\lambda_a \le 1$), 
due to the spin-chirality driven mechanism. 
In the opposite case $\rho_a \gg \rho_c$ the transverse current appears due to spin Hall effect 
induced by the adiabatic Berry phase mechanism.
This regime is set when 
the spin-flip channels are suppressed at  $\lambda_a \gg 1$. 
With the change of the texture diameter $a$ from small to large values
a crossover occurs between the charge Hall and spin Hall dominated regimes. 
In addition, the change of the texture size affects the wave parameter $ka$, which determines the properties of the scattering related to the electron relation between the electron wavelength and the scatterer size. 
With these two consequences of the texture size variation acting simultaneously, the topological Hall resistivity $\rho_{yx}^T$ necessarily exhibits a non-monotonic dependence on $ka$.

Fig.~\ref{f1}(b) shows the calculated dependence of 
charge $\rho_c$, adiabatic $\rho_a$ and total $\rho_{yx}^T$ topological Hall resistivity on the skyrmion diameter $a$ for the magnetic skyrmion with magnetization spatial profile $\Lambda_1(r)$ shown in  Fig.~\ref{f3}.
For the calculation results shown in Fig.~\ref{f1} the spin polarization was taken $P_s = 0.4$, and the skyrmion sheet density $n_{sk} = 2 \times 10^{11}$ cm$^{-2}$.
The scattering rates $\mathcal{J}_{ss'}$ were calculated using the phase function method~\cite{SciRep_skyrmion}.

As can be seen in Fig.~\ref{f1}, for $\lambda_a \le 1.8$ 
the charge 
contribution $\rho_c$ exceeds $\rho_a$, at that 
$\rho_{yx}^T$ is dominated by the purely charge current. 
For $\lambda_a \ge 4.5$ the adiabatic 
term 
prevails 
$\rho_a \gg \rho_c$
and $\rho_{yx}^T$ appears due to the spin current converted into the charge current. 
As was discussed above, indeed  $\rho_{yx}^T$ appears to be non-monotonic when the crossover between $\rho_c$ and $\rho_a$ occurs in the range ($ 1.8 \gtrsim \lambda_a \gtrsim 4.5 $).
As the  spin-flip processes are suppressed, 
firstly the charge contribution $\rho_c$ is decreased, and only later the adiabatic term $\rho_a$
starts to increase. 
This effect results in the appearance of local minimum for $\rho_{yx}^T$ in the crossover regime. 



\begin{figure}
	\centering	
	\includegraphics[width=0.5\textwidth]{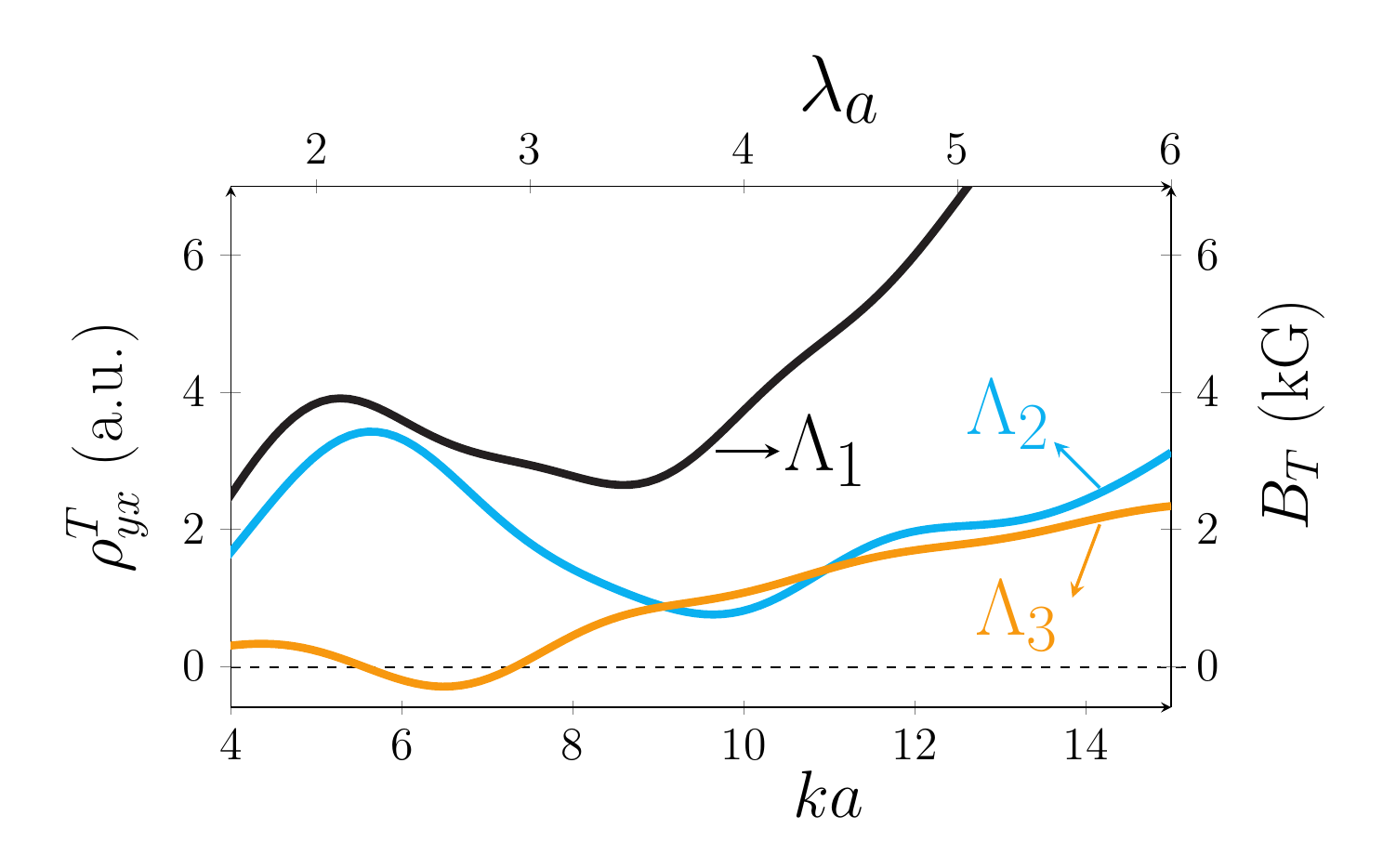}
	\caption{The dependence of $\rho_{yx}^T$ on chiral texture diameter $ka$ for different texture profiles in the region of crossover. The parameters $P_s =0.4$, $n_{sk} = 2\times 10^{11}$ cm$^{-2}$.}	
	\label{f5}
\end{figure}

The behaviour of $\rho_{yx}^T$ in the crossover regime is highly sensitive to a particular magnetic texture profile. 
In Fig.~\ref{f5} we present the dependence of $\rho_{yx}^T$ on $a$ for three different spin texture spatial profiles shown in Fig.~\ref{f3}.
As can be seen in Fig.~\ref{f5} 
the oscillating structure of $\rho_{yx}^T$ upon increasing $ka$ exhibits a significant variation even for two very similar skyrmion configurations $\Lambda_1$ and $\Lambda_2$. 
The strong dependence of $\rho_{yx}^T$ on $\Lambda(r)$ observed in the crossover regime is due to 
the significance of the interference in electron scattering as the wave parameter $ka\sim\pi$ (Ref.~\cite{SciRep_skyrmion}). 
The texture described by $\Lambda_2$ has larger spin gradients, so the adiabatic term activates at larger $ka$, and the magnitude of $\rho_{yx}^T$ for $\Lambda_2$ in the crossover regime is smaller than that of $\Lambda_1$.

We would also like to stress out that the topological Hall effect exists as well 
due to scattering on non-collinear spin rings having zero winding number (orange curve in Fig.~\ref{f5}). 
The transverse conductivity $\rho_{yx}^T$ due to scattering on non-collinear spin rings possesses all the  features described above including the the existence of charge and spin Hall limiting regimes. 

\subsubsection{The magnitude of the topological Hall effect}

The magnitude of THE for the dilute systems can be expressed in terms of the effective magnetic field $B_T$  
introduced as:
\begin{equation}
    \label{eq_BT}
    \begin{aligned}
\rho_{yx}^T = \frac{B_T}{nec}. 
    \end{aligned}
\end{equation}
The field $B_T$ shows the 
 magnitude of the  external magnetic field 
applied to the sample, at which the ordinary Hall effect contribution to the transverse resistivity $\rho_{yx}^{\mathcal{O}}$ becomes comparable with $\rho_{yx}^T$.


Usually in the THE estimates it is assumed that each skyrmion contributes
via a magnetic flux quantum~\cite{DiscretHall}. 
However, our analysis shows that 
such an estimate does not take 
into account the 
important features of the scattering. 
According to 
Eqs.~(\ref{eq_rhoT}),(\ref{eq_BT}), $\rho_{yx}^T$ and $B_T$ linearly depend 
on both the skyrmion sheet density $n_{sk}$ and the dimensionless scattering rates $\mathcal{J}_{ss'}$. 
Therefore, the actual magnitude of $B_T$ is renormalized 
differently depending on the scattering regime.
The scattering rates $\mathcal{J}_{ss'}$ are small at $\lambda_a \le 1$, of the order of unity at an intermediate $\lambda_a\sim1$, and reach the order of tens at  $\lambda_a \gg 1$.

\begin{figure}
	\centering	
	\includegraphics[width=0.5\textwidth]{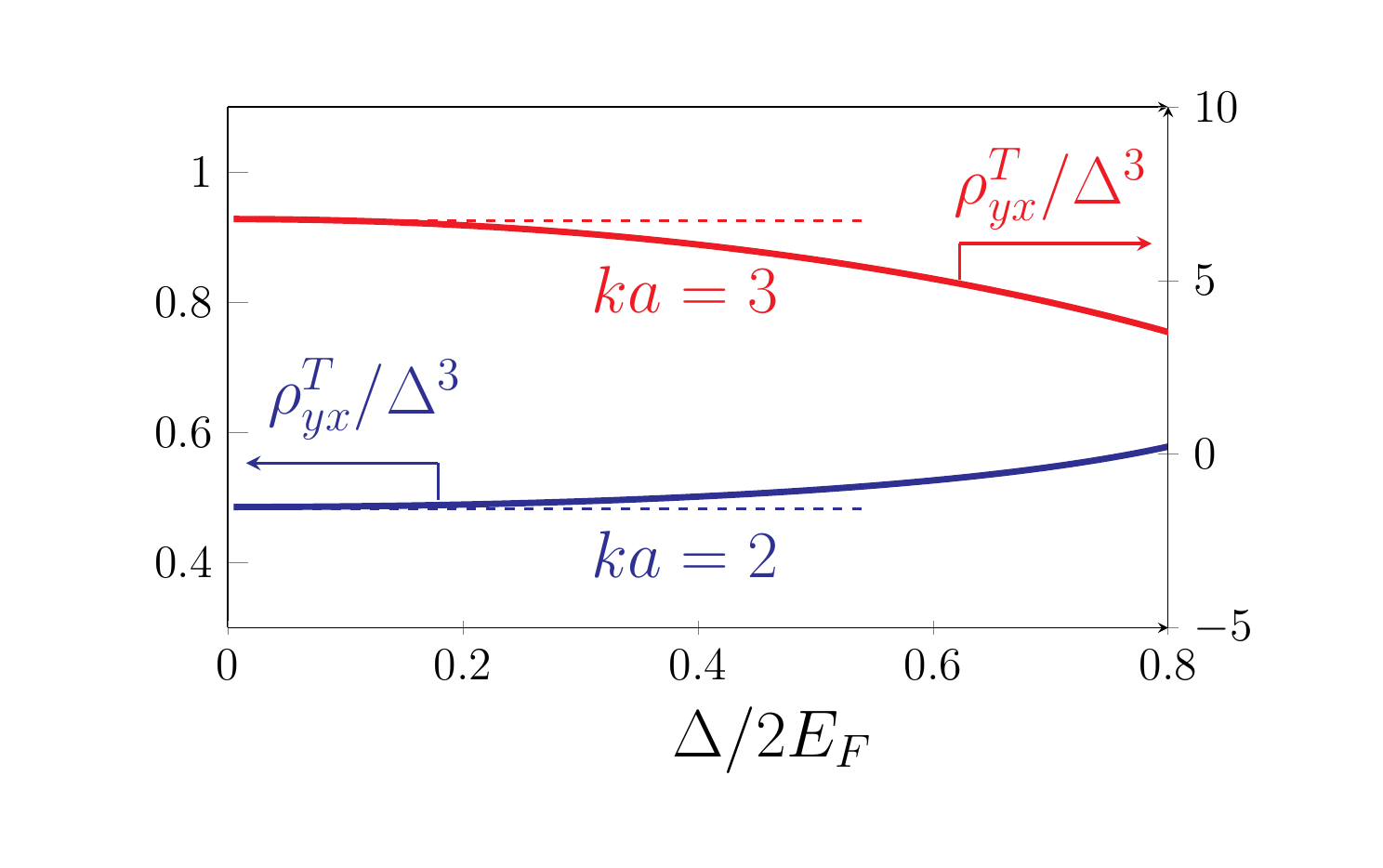}
	\caption{The dependence of $\rho_{yx}^T/\Delta^3$ on the variation of $\Delta/2E_F$ at $ka=2$ (blue curve) and $ka=3$ (red curve) for $\Lambda_1$ profile.}	
	\label{fd3}
\end{figure}

The magnitude of $B_T$ for $n_{sk} = 2 \times 10^{11}$ cm$^{-2}$
can be seen in Figs.~\ref{f1}-\ref{f2-old}.
At $n_{sk} \phi_0 \approx 8$ T the value of $B_T$ in the intermediate regime is of the order of several kG; in the strong coupling regime ($\lambda_a \gg 1$) it  
can go as high as several Tesla~\cite{denisov2018topological}. 

In the weak coupling regime $\rho_{yx}^T$ scales as $\Delta^3$, 
as the perturbation theory couples $\rho_{yx}^T$ with spin chirality and, therefore, 
THE requires the third order in the exchange interaction. 
In Fig.~\ref{fd3} the quantity $\rho_{yx}^T/ \Delta^3$ is shown for two values of $ka=2,3$.
As can be seen from the figure, 
the scaling $\Delta^3$ holds up to $\Delta/2 E_F \approx 0.2$, the 
 deviation from the scaling relation indicates 
 that 
 the perturbation theory becomes invalid departing from the weak coupling regime. 

Let us note, that although the asymmetrical scattering rates $\mathcal{J}_{ss'}$ are small in the weak coupling regime due to the strong dependence on the exchange interaction strength $\Delta$ ($\mathcal{J}_{ss'}$ is proportional to $(\Delta/2E_F)^3 (ka)^8$ at $\lambda_a \ll 1$), the magnitude of $B_T$ can be large due to higher spin textures sheet density. For example, for $n_{sk} = 5 \times 10^{12}$ cm$^{-2}$ and $\lambda_a =0.8$ ($ka = 2$, $P_s = 0.4$) one gets $B_T \approx 0.7$ T. 

\subsubsection{The sign of the topological Hall effect}

In a real experiment when electron transport in a system with non-collinear spin textures 
is studied 
as a function of the external magnetic field $B_0$, it is often difficult to extract different contributions to the Hall effect.
The total transverse resistivity $\rho_{yx}$ contains three contributions $\rho_{yx} = \rho_{yx}^{\mathcal{O}} + \rho_{yx}^{\mathcal{A}} + \rho_{yx}^T$, where $\rho_{yx}^{\mathcal{O}},\rho_{yx}^{\mathcal{A}}, \rho_{yx}^{T}$ are attributed to the ordinary, anomalous and topological Hall effects respectively. 
Here we focus on the sign difference  
between $\rho_{yx}^T = (B_T/nec)$ and $\rho_{yx}^{\mathcal{O}} = (B_0/nec)$, 
thus we should compare the signs of $B_T$ and $B_0$.


We assume, that the background magnetization $S_0$ is directed along the external magnetic field $B_{0}>0$. 
In general, there is now any fixed relation between the signs of the topological Hall resistivity $\rho_{yx}^T$, charge $\rho_c$ and adiabatic $\rho_a$ contributions
as can be seen in 
Fig.~\ref{f1} and Fig.~\ref{f5}. In these figures $\rho_a$ and $\rho_{yx}^T$ for $\Lambda_3$ spin configuration 
change their signs  upon increase of $ka$ in the crossover region. 
However, it is possible to specify the sign of $B_T$ 
in the limiting regimes, i.e. away from the threshold $E_F \gg \Delta/2$ and outside the adiabatic crossover $\lambda_a \approx 1$. 
Let us consider the weak coupling regime ($\lambda_a \le 1$), in which the charge current contribution to THE  dominates ($\rho_c \gg \rho_a$).
In this regime, the effective magnetic field is proportional to the chirality of the spin texture 
$B_T \propto \delta \boldsymbol{S}_1 \cdot \left[ \delta \boldsymbol{S}_2 \times \delta \boldsymbol{S}_3 \right]$.
For $\varkappa = +1$, $\eta = +1$ the sign of the mixed vector product of any three spins $\delta \boldsymbol{S}_1,\delta \boldsymbol{S}_2,\delta \boldsymbol{S}_3$ forming the skyrmion
 is negative and $B_T<0$ due to $\delta S_z <0$ (see Fig.~\ref{f3}), thus the sign of   
$B_T$ appears to be opposite to $B_0$.
In the adiabatic regime ($\rho_a \gg \rho_c$) the electrons with positive spin projection (co-aligned with $S_0$) retain the same type of scattering asymmetry as for  small $\lambda_a$. 
As these electrons constitute the majority at a  positive spin polarization ($P_s>0$),
the effective magnetic field is also negative $B_T<0$. 

We conclude, that for $\varkappa =+1$, $\eta=+1$ configurations, the topological field $B_T$ 
usually has 
the opposite sign to the sign of the external field $B_0$. 
For chiral spin configurations with negative vorticity $\varkappa<0$ the fields $B_T$ and $B_0$ have the same sign. 
However, 
in the crossover regime $\lambda_a \sim 1$ and near the threshold $E_F \approx \Delta/2$ 
there is no any fixed relation between $B_0$
and $B_T$ signs. 





\subsubsection{Effect of the Fermi energy variation}

\begin{figure}
	\centering	
	\includegraphics[width=0.5\textwidth]{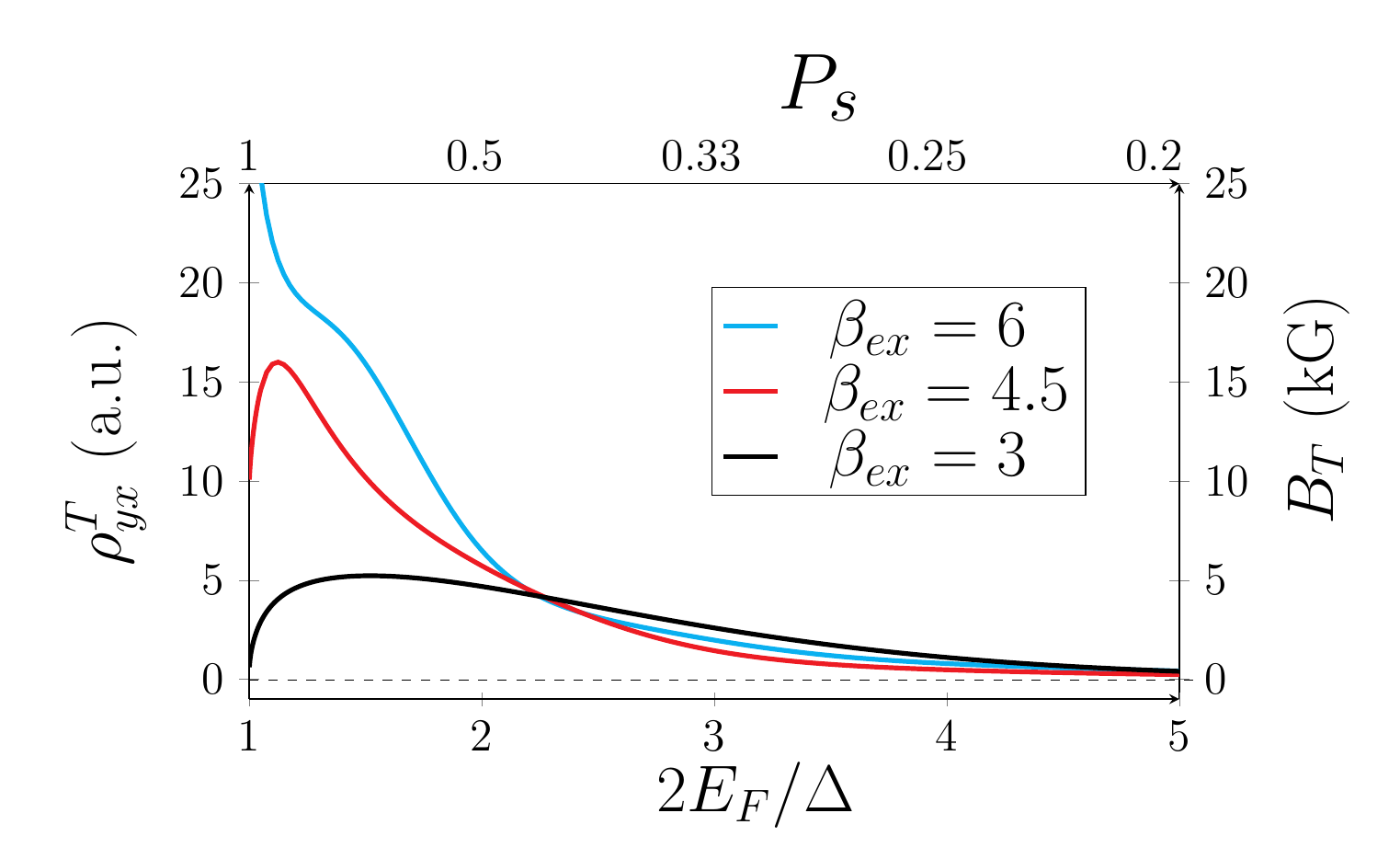}
	\caption{The dependence of $\rho_{yx}^T$ on Fermi energy $E_F$ at different $\beta_{ex} = \sqrt{m \Delta/\hbar^2} a$ parameter for $\Lambda_1$ profile.}	
	\label{f2}
\end{figure}

\begin{figure}
	\centering	
	\includegraphics[width=0.5\textwidth]{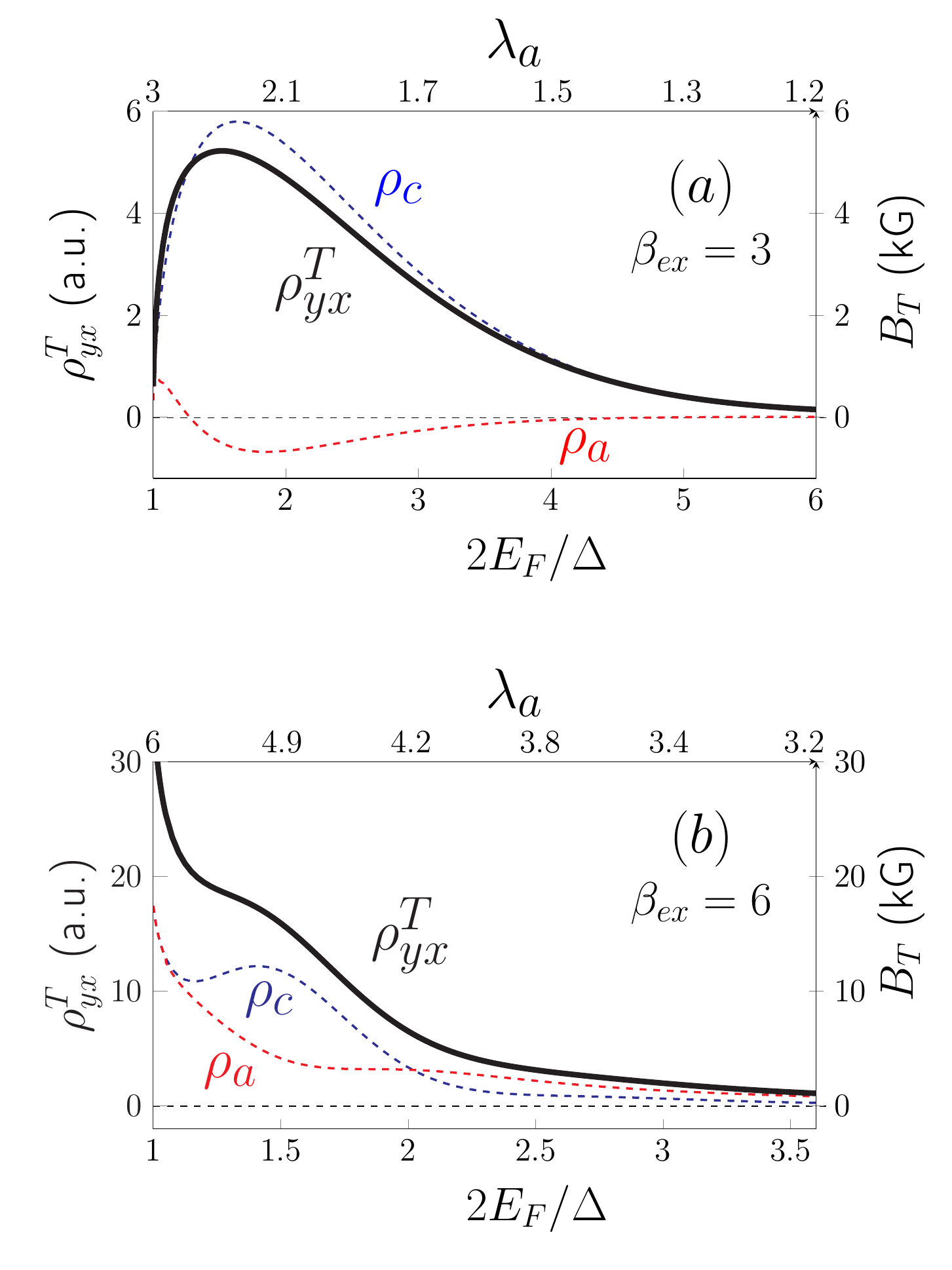}
	\caption{The dependence of $\rho_{yx}^T$, $\rho_c, \rho_a$ on Fermi energy $E_F$ at $\beta_{ex} = 3$ (a) and $\beta_{ex}=6$ (b) for $\Lambda_1$ profile.}	
	\label{f2-old}
\end{figure}

The dependence of $\rho_{yx}^T$ on the variation of the Fermi energy $E_F$ exhibits a number of distinctive features. 
At $E_F < \Delta/2$ only one spin subband is occupied and  spin polarization is $P_s=1$. 
We start the analysis 
from the threshold $E_F \ge \Delta/2$, when 
the electrons start populating the second spin subband.  
In further consideration we keep $\Delta$ and $a$ constant changing only the Fermi energy $E_F$, which we express though the dimensionless parameter $2 E_F/\Delta = P_s^{-1}$. 
We also introduce a dimensionless spin texture diameter $\beta_{ex} = \sqrt{m \Delta/\hbar^2} a$, it 
is independent of $E_F$.   
Fig.~\ref{f2} shows the dependence of $\rho_{yx}^T$ on $2E_F/\Delta$ and $P_s$ calculated for the $\Lambda_1$ skyrmion configuration for three different values of $\beta_{ex}$. 
As can be seen from the figure, 
$\rho_{yx}^T$ depends non-monotonically on $2E_F/\Delta$, with a maximum near the threshold and decreasing at a larger $E_F$. 
The suppression of $\rho_{yx}^T$ at a large $E_F$ is due to  destructuve scattering interference at $P_s \ll 1$ and $ka \gg 1$ (see Ref.\cite{SciRep_skyrmion}). 
The magnitude of $\rho_{yx}^T$ near the threshold is controlled by $\beta_{ex}$. As the spin-chirality driven mechanism relevant for a small skyrmion size does not work at $E_F < \Delta/2$ (there is no spin flip processes below the spin down subband edge), the decrease of $\beta_{ex}$ suppresses $\rho_{yx}^T$ at $E_F=\Delta/2$. 
These features of $\rho_{yx}^T$ are specific to THE and 
can be possibly used to distinguish it from AHE and OHE contributions. 



The variation of the Fermi energy 
affects the
asymmetric part of the scattering cross-section 
simultaneously through $ka$ and $P_s$ factors and, therefore,
gives rise to a number of interesting features in the transverse resistivity  
behaviour. 
We demonstrate these peculiarities in Fig.~\ref{f2-old}, where the dependence of $\rho_{yx}^T$, $\rho_c$, and $\rho_a$ on $2E_F/\Delta$ is plotted for $\beta_{ex} = 3$  and $\beta_{ex} = 6$. 
The variation of $E_F$ directly affects 
the adiabatic parameter, which can be expressed as  
$\lambda_a = \beta_{ex} \sqrt{P_s}$.
For $\beta_{ex} = 3$ (Fig.~\ref{f2-old}a) the adiabatic term $\rho_a$ is negative when far from the threshold. This is due to the complex scattering pattern 
typical for the 
intermediate range of the adiabatic parameter values ($1 \le \lambda_a \le 2$). 
We have already encountered this effect considering the behaviour of THE in the crossover regime: 
$\rho_a$ is negative in Fig.~\ref{f1} for the same range of $\lambda_a$ as in Fig.~\ref{f2-old}a.
For $\beta_{ex} = 6$ (Fig.~\ref{f2-old}b) $\lambda_a$ is larger and the interference in the carrier scattering 
manifests itself through the oscillation of $\rho_c,\rho_a$ magnitudes superimposed on the global suppression upon increasing of $E_F$.
The same oscillating peculiarities of transverse response can be seen in Fig.~\ref{f1} in the range  $4 \le \lambda_a \le 5$. 
Let us finally comment on the scattering rates behaviour
in the vicinity of the threshold $E_F \approx \Delta/2$. 
Since the spin down and spin flip scattering channels are absent below the threshold 
$E_F< \Delta/2$, we conclude that at $E_F \approx \Delta/2$ the following relations are fulfilled: $\varGamma_2 \approx 0$, and $\varGamma_1(\theta) \approx \Pi(\theta)$. 
At that, only spin up scattering channel is activated with $\mathcal{J}_{\uparrow \uparrow} \approx 2 \Pi(\theta)$ (i.e. $\rho_a \approx \rho_c$).

\subsection{Dense array of skyrmions}
\label{sb_trans}

In this section we apply the diffusive theory of THE 
to the case 
when the dominating scattering mechanism changes from scattering on non-magnetic impurities to scattering on magnetic textures. 
This transition takes place, for example, in ferromagnetic films in the vicinity of phase transition~\cite{Moca2009}, when the sheet density of thermally activated chiral magnetic fluctuations increases~\cite{kamal1993new,AronzonRozh,Ye1999}.  
The exchange interaction between free carriers and localized magnetic moments is typically strong in these systems so that the adiabatic approximation for THE is applicable. 
At that,
the spin-flip scattering channels are completely suppressed ($\Omega_{\uparrow \downarrow} = \tau_{\uparrow \downarrow}^{-1} =0$), and THE originates solely from the spin Hall effect ($\rho_a \gg \rho_c$). 




The switching of the dominant scattering mechanism affects the 
spin dependent scattering time $\tau_s$ (Eq.~\ref{eq_rate}).
In the dilute regime of low skyrmion density 
$ \omega_s \tau_0  \ll 1$ 
considered in the previous section, 
the total transport scattering time $\tau_s$ is independent 
%
of the carrier spin, being 
determined by scattering on host non-magnetic impurities $\tau_s = \tau_0$. 
Increasing the skyrmion sheet density 
turns the system into 
the dense skyrmionic regime $ \omega_s \tau_0 \gg 1$, when the total transport scattering time is determined solely by the magnetic skyrmions
and, hence, depends on the carrier spin state 
$\tau_s = \omega_{s}^{-1}$.
This transition 
affects both the longitudinal 
and transverse 
resistivities. 
Even in the dense skyrmionic regime the magnitude of $ \Omega_{ss} \tau_{s} \approx \Omega_{ss}/\omega_{s} $ remains small, as the symmetric scattering is more effective than the asymmetric one~\cite{denisov2018topological}. 
Thus, we can still solve the kinetic equation in the lowest order in $ \Omega_{ss}\tau_{s}$ as described in the previous section. 

\begin{figure}
	\centering	
	\includegraphics[width=0.5\textwidth]{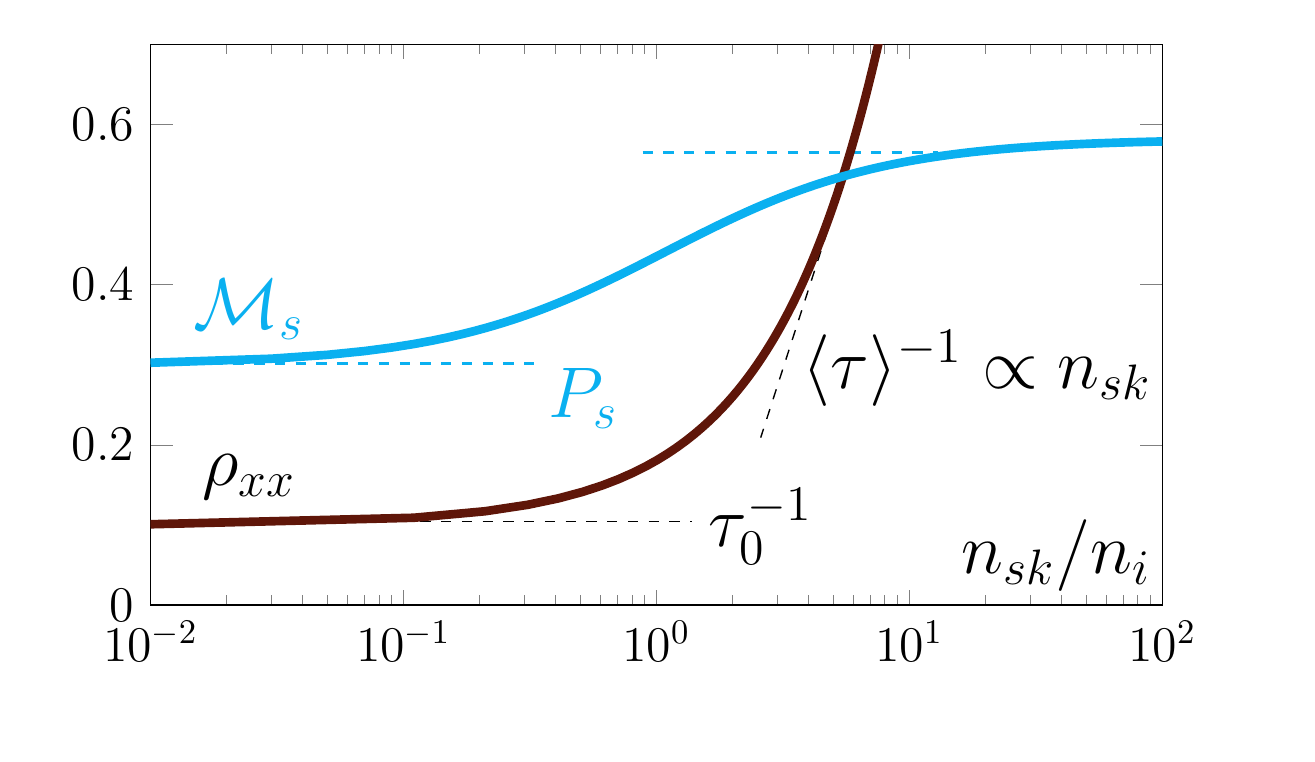}
	\caption{The dependence of $\rho_{xx}$ and $\mathcal{M}_s$ on skyrmion sheet density $n_{sk}$.}	
	\label{f7}
\end{figure}

Since the spin-flip processes are suppressed {in the adiabatic regime}, 
the spin up and spin down channels are uncoupled and contribute independently to the conductivity. Keeping only the leading terms with respect to $\Omega_{ss}\tau_{s}$ we get for the 
resistivity tensor:
\begin{equation}
\label{eq_res1}
    \begin{aligned}
    & \rho_{xx} = \frac{m}{n e^2 \langle \tau \rangle},
    \hspace{0.6cm}
     \rho_{yx}^T = \mathcal{M}_s 
     \frac{1}{nec} \left(\phi_0 n_{sk}\right) \int\limits_0^{2\pi} \Pi \sin{\theta} d\theta,
    \\
    & \langle \tau \rangle = \frac{1}{2} \left[  (1+P_s) \tau_{\uparrow} + (1-P_s) \tau_{\downarrow}
    \right],
    \\
&\mathcal{M}_s = \frac{1}{2} \left[ (1+P_s) \frac{\tau_{\uparrow}^2}{\langle \tau \rangle^2} -  (1-P_s) \frac{\tau_{\downarrow}^2}{\langle \tau \rangle^2} \right].
\end{aligned}
\end{equation}
Here $P_s = (n_{\uparrow}-n_{\downarrow})/(n_{\uparrow}+n_{\downarrow})$ is the spin polarization
of the 2D free carriers.
We have also introduced an averaged scattering time $\langle \tau \rangle$.
The introduced parameter $\mathcal{M}_s$ controls the conversion of the spin Hall to the charge Hall current.

In Fig.~\ref{f7} we plot the dependence of $\rho_{xx}$ and the spin/charge Hall factor $\mathcal{M}_s$ on the skyrmion sheet density $n_{sk}$ via the parameter $\omega_s \tau$ covering the transition between scattering on non-magnetic impurities and skyrmions.
In the dilute regime $\langle \tau \rangle = \tau_0$, and $\rho_{xx}$ does not depend on $n_{sk}$.
When $\tau_{ss}^{-1}$ exceeds $\tau_0$,
the longitudinal resistivity $\rho_{xx} \propto \langle \tau \rangle^{-1}$
increases linearly with $n_{sk}$ 
as shown  in Fig.~\ref{f7}. 
From the experimental point of view this transition 
leads to the peak in resistivity 
upon increase of the temperature in the vicinity of the phase transition. 



According to Eq.~(\ref{eq_res1}) the topological Hall resistivity $\rho_{yx}^T$ is proportional to the skyrmion sheet density $n_{sk}$.
The crossover in the dominating scattering mechanism affects $\rho_{yx}^T$ only via the $\mathcal{M}_s$-factor. 
In the dilute regime ($\omega_s \tau_0 \ll 1$) this parameter coincides with the carrier spin polarization $\mathcal{M}_s = P_s$ as the scattering time on host impurities $\tau_0$ is spin independent. 
However, in the dense regime  ($\omega_s \tau_0 \gg 1$) the scattering time $\tau_s$ depends on the carrier spin,  
this dependence  creates an additional spin imbalance favoring the conversion of spin to charge currents. 
As a result, the $\mathcal{M}_s$-factor is renormalized accounting for $\tau_{\uparrow} \neq \tau_{\downarrow}$.


Let us mention, that to describe dependence of the resistivities $\rho_{xx}, \rho_{yx}$ 
on temperature and external magnetic field in the vicinity of FM-transition 
one should specify a specific model of skyrmion/antiskyrmion creation 
adequate for the considered material system. 


The general expressions for $\rho_{xx}$ and $\rho_{yx}^T$ in the adiabatic regime (\ref{eq_res1}) are applicable for 
any spin-dependent scattering mechanisms, not necessarily due to skyrmions. 
We point out, that in the leading order with respect to $\Omega_{ss} \tau_s$ the effect of $\tau_s$ on $\rho_{yx}^T$ can be fully decribed by the replacement of the carriers spin polarization $P_s$ by an effective $\mathcal{M}_s$-factor, which accounts for  $\tau_{\uparrow} \neq \tau_{\downarrow}$.

\subsection{Paramagnetic chiral systems}
\label{sb_semicond}




In the previous sections we considered THE in a 2D magnetic layer with a
background magnetization $S_0$ and local deviations forming chiral magnetic textures. 
Unlike anomalous Hall effect, THE does not necessarily require macroscopic spin polarization of the carriers in the sample.
Therefore, THE is allowed in a system with no background magnetization provided it still has localized  chiral spin textures.
We will refer to this situation as to a chiral paramagnetic case. 
In the absence of a preferred magnetization direction the chiral spin 
textures with opposite orientations can be created in the same sample.
A few special of features of $\rho_{yx}^T$ are expected in such a case. 

\begin{figure}
	\centering	
	\includegraphics[width=0.5\textwidth]{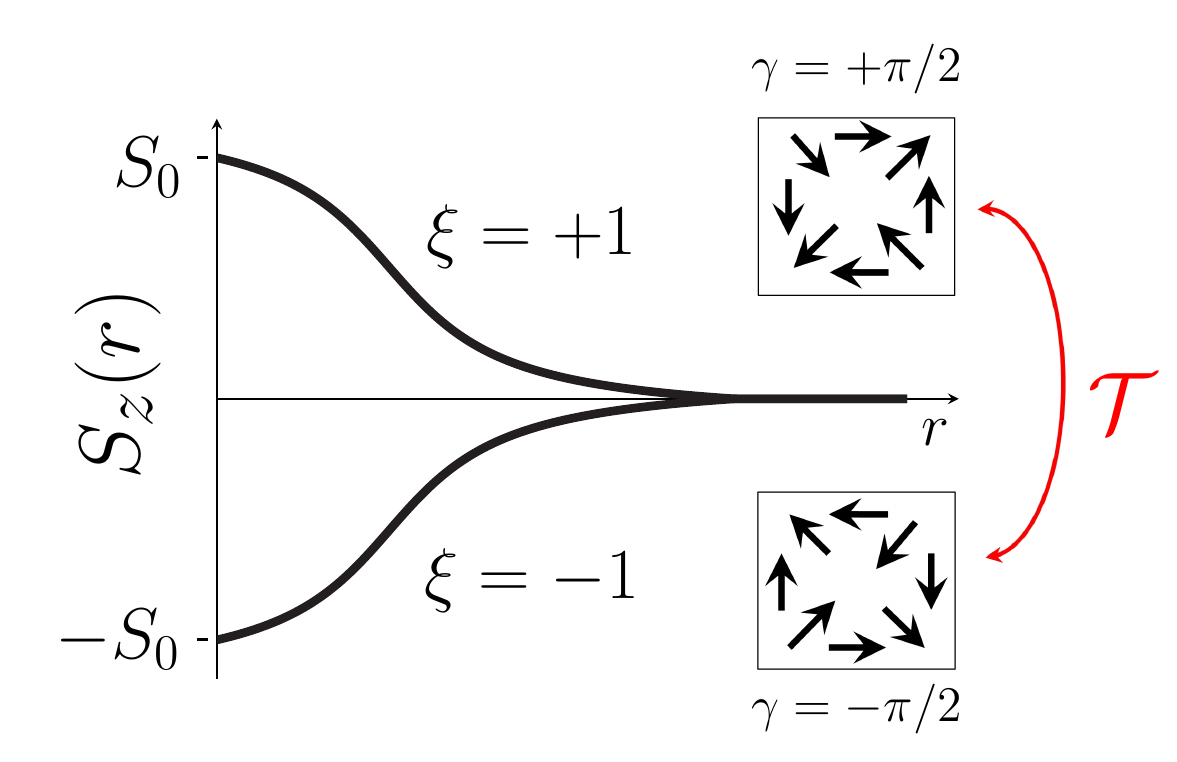}
	\caption{Two chiral spin textures with opposite orientation $\xi = \pm 1$ and $\varkappa = -1$ connected by the time-inversion $\mathcal{T}$.}
	\label{f9}
\end{figure}

One scenario leading to a chiral paramagnetic system 
%
%
%
with two independent spin orientations 
has been recently proposed on the basis of a magnetic polaron (BMP) in a semiconductor~\cite{denisov2018hall}.
BMP is a collective state of a carrier localized at an impurity center and other paramagnetic impurities lying within the localization radius.  
Due to the exchange interaction 
the carrier 
polarizes other magnetic impurities within its wavefunction radius;  
at that a non-collinear spin strucure 
of a chiral BMP 
appears due to the spin-orbit splitting of the carrier band states. 
This effect 
can be 
viewed as a case of Dzyaloshinskii-Moriya interaction mediated by a localized electron state.    
There are two opposite orientations of the spin field forming chiral BMP 
 connected by time inversion symmetry.
We denote these two configurations by the orientation of spins in the center $\xi = {\rm sgn}(S_z|_{r\to 0}) = \pm 1$. 
The example of the doublet of chiral BMPs with account for Dresselhauss spin-orbit interaction is shown in Fig.~\ref{f9}.

The presence of two spin textures with opposite orientation $\xi = \pm 1$ in the same layer 
modifies the expression for the charge contribution $\rho_c$ to THE (\ref{eq_rhoT}).
Indeed, as $S_0 \approx 0$ 
the sign of the spin-chirality driven contributions $\varGamma_{1,2}$ 
to the carrier asymmetric scattering on the spin texture depends on its orientation
and the contributions to  $\rho_c$ 
from textures with $\xi = \pm 1$  have opposite sign.
We arrive at the modified expression for $\rho_c$ accounting for both 
texture orientations $\xi = \pm 1$:
\begin{equation}
    \label{eq_rc_BMP}
    \begin{aligned}
    &\rho_{c} = P_{\xi} \frac{1}{nec} (\phi_0 n_{sk}) \int (\varGamma_1 + \varGamma_2) \sin{\theta} d\theta,
    \\
    & P_{\xi} = \frac{n_+ - n_-}{n_{+}+n_-},
    \end{aligned}
\end{equation}
where $n_\pm$ are  sheet densities of $\xi = \pm 1$ spin textures, respectively,
$n_{sk} = n_+ + n_-$ is the total sheet density, $P_{\xi}$ is the polarization of the texture array in terms of their orientations.  
Here we consider the dilute regime with
$ \tau_s = \tau_0$. 
It follows from (\ref{eq_rc_BMP}), that  observation of THE in chiral  paramagnetic systems is possible only when there is 
an imbalance in the texture orientations i.e.  
$P_{\xi} \neq 0$. 
Let us note, that for the positive texture polarization $P_{\xi} >0$, the sign of $\rho_c$ is different to that of $\rho_c$ for magnetic skyrmions, or non-collinear rings in Fig.~\ref{f3}. 
Indeed, as we already mentioned, $\delta S_z <0$  for the magnetic skyrmions case leading to $\rho_c >0$. 
On the contrary, $\delta S_z >0$ is positive for $\xi = +1$ shown in Fig.~\ref{f9}, so that $\rho_c <0$.




\section{Summary}

\label{s_5}

We have developed a theory of the topological Hall effect in a 2D system with randomly located  chiral magnetic textures. 
We calculated THE resistivity $\rho_{yx}^T$ on the basis of Boltzmann kinetic equation 
accounting for 
the carrier scattering asymmetry on a localized chiral spin texture. 
We have shown, that 
$\rho_{yx}^T$ 
can be expressed as a sum of two contributions: $\rho_{yx}^T = \rho_c + \rho_a$. 
The first one $\rho_c$ describes 
the transverse charge current due to spin-independent asymmetric scattering, 
while the second one $\rho_a$ is spin-dependent and 
describes the spin Hall effect, contributing in its turn 
to the charge Hall current provided the free carriers are spin polarized. 
We have investigated the interplay between 
adiabatic and non-adiabatic regimes of free carriers scattering on spin textures. We 
predict the appearance of a local minimum in the dependence of $\rho_{yx}^T$ on a skyrmion size 
associated with the crossover from charge Hall dominating to spin Hall dominating regime. 
The non-monotonic features were also found for the dependence of THE on 
the carriers Fermi energy. 
We have obtained general expressions for longitudonal and transverse components of the resistivity tensor 
upon the transition from dilute to dense skyrmion array in the particularly important case of the adiabatic scattering. 
Finally, we have clarified the role of spin-independent charge contribution to THE 
in dilute magnetic systems with zero background magnetization, showing that the sign of $\rho_c$ 
is determined by the orientation of chiral spin texture in its center. 






\section{Acknowledgments}
We thank V.~Cross, A.~Fert, D.~Maccariello, N.~Reyren for a very fruitful discussions and comments.

The work has been carried out under the financial support of
Grants from the Russian Science Foundation (asymmetric scattering theory - project no.~17-12-01182; numerical calculations - project no.~17-12-01265) 
and from Russian Foundation of Basic Research (grant 18-02-00668).
It was also supported by the Academy of Finland Grant No.~318500.
K.S.D. and N.S.A. thank the Foundation for the Advancement of Theoretical Physics and Mathematics "BASIS".

\appendix

\section{Symmetry of scattering rates}

\label{Ap1}

In this Appendix we derive the relations (\ref{eq_j}) for the dimensionless asymmetric scattering rates $\mathcal{J}_{ss'}(\theta,\eta)$. 
The background polarization 
$\eta = {\rm sgn}\left(S_0\right) = \pm 1$, it also  
determines the orientation of spins inside the core of a chiral spin texture. 
The starting point is the time-reversal invariance, which states that 
if we make a replacement $\boldsymbol{S}(\boldsymbol{r}) \to -\boldsymbol{S}(\boldsymbol{r})$, 
then 
the scattering rate from $(\boldsymbol{p}',s') \to (\boldsymbol{p},s)$ 
with a scattering angle $\theta = \varphi - \varphi'$ 
is equal to that of $(-\boldsymbol{p},\bar{s}) \to (-\boldsymbol{p}', \bar{s}')$ 
with the scattering angle $-\theta$ 
($\bar{s}$ denotes the carrier spin state opposite to $s$). 
The replacement $\boldsymbol{S}(\boldsymbol{r}) \to -\boldsymbol{S}(\boldsymbol{r})$ leads to 
$\eta \to -\eta$, $\varkappa \to \varkappa$, and $\gamma \to \gamma + \pi$. 
Collecting these operations together  
we obtain: 
\begin{equation}
    \label{eq_a1}
    \mathcal{G}_{ss'}(\theta,\eta) + \mathcal{J}_{ss'}(\theta,\eta) = 
    \mathcal{G}_{\bar{s}' \bar{s}}(-\theta,-\eta) + \mathcal{J}_{\bar{s}' \bar{s}}(-\theta,-\eta) 
\end{equation}

Taking into account that $ \mathcal{G}_{ss'}(\theta,\eta) = \mathcal{G}_{ss'}(-\theta,\eta)$ and $\mathcal{J}_{ss'}(\theta,\eta) = - \mathcal{J}_{ss'}(-\theta,\eta) $ 
we get that symmetric $\mathcal{G}_{ss'}$ and asymmetric $\mathcal{J}_{ss'}$ rates should satisfy:
\begin{equation}
    \begin{aligned}
    & \mathcal{G}_{ss'}(\theta,\eta)  = \mathcal{G}_{\bar{s}' \bar{s}}(\theta,-\eta) ,
    \\
    & \mathcal{J}_{ss'}(\theta,\eta) = -  \mathcal{J}_{\bar{s}' \bar{s}}(\theta,-\eta). 
    \end{aligned}
    \label{eq_ap_r}
\end{equation}
We further focus on $ \mathcal{J}_{ss'} $. 
The relations (\ref{eq_ap_r}) couple the two scattering channels with the opposite spin orientations. 
For the spin-conserving channels we have:
\begin{equation}
    \mathcal{J}_{\uparrow \uparrow}(\theta,\eta) = -  \mathcal{J}_{\downarrow \downarrow}(\theta,-\eta). 
    \label{eq_Jcons}
\end{equation}
Let us introduce the symmetrized and antisymmetrized combinations of $\mathcal{J}_{\uparrow \uparrow}(\theta,\eta)$, $\mathcal{J}_{\uparrow \uparrow}(\theta,-\eta)$ with respect to $\eta$: 
\begin{equation}
    \begin{aligned}
    & \varGamma(\theta,\eta) = \frac{1}{2}\left( \mathcal{J}_{\uparrow \uparrow}(\theta,\eta) - \mathcal{J}_{\uparrow \uparrow}(\theta,-\eta) \right),
    \\
    & \Pi(\theta) = \frac{1}{2}\left( \mathcal{J}_{\uparrow \uparrow}(\theta,\eta) + \mathcal{J}_{\uparrow \uparrow}(\theta,-\eta) \right).
    \end{aligned}
\end{equation}
Since the background polarization  $\eta = \pm 1$ can take only two values, it is obvious that 
the function $\Pi$ does not depend on $\eta$, while $\varGamma(-\eta) = - \varGamma(\eta)$.  
Let us extract the dependence of $\varGamma$ on $\eta$ in the explicit form: $\varGamma(\eta,\theta) \equiv \eta \varGamma_1(\theta)$, where $\varGamma_1$ depends only on $\theta$ and on the energy of an incident electron. 
Expressing the rates of the spin-conserving channels and using the symmetry (\ref{eq_Jcons}) we arrive at the relations (\ref{eq_j}):
\begin{equation}
    \begin{aligned}
    & \mathcal{J}_{\uparrow \uparrow}(\theta,\eta) = \eta \varGamma_1(\theta) + \Pi(\theta),
    \\
    & \mathcal{J}_{\downarrow \downarrow}(\theta,\eta) = \eta \varGamma_1(\theta) - \Pi(\theta).
    \end{aligned}
\end{equation}


As for the spin-flip channels, there is an additional symmetry $\mathcal{J}_{\uparrow \downarrow}(\theta,\eta) = \mathcal{J}_{\downarrow \uparrow}(\theta,\eta)$  as the  Hamiltonian is hermitian, this symmetry leads to the absence of a spin Hall part:
\begin{equation}
    \label{eq_a3}
    \mathcal{J}_{\downarrow \uparrow}(\theta,\eta) = \mathcal{J}_{\uparrow \downarrow}(\theta,\eta) = \eta \varGamma_2(\theta), 
\end{equation}
where $\varGamma_2(\theta)$ does not depend on $\eta$.


\bibliography{Skyrmion}

\begin{thebibliography}{53}
\expandafter\ifx\csname natexlab\endcsname\relax\def\natexlab#1{#1}\fi
\expandafter\ifx\csname bibnamefont\endcsname\relax
  \def\bibnamefont#1{#1}\fi
\expandafter\ifx\csname bibfnamefont\endcsname\relax
  \def\bibfnamefont#1{#1}\fi
\expandafter\ifx\csname citenamefont\endcsname\relax
  \def\citenamefont#1{#1}\fi
\expandafter\ifx\csname url\endcsname\relax
  \def\url#1{\texttt{#1}}\fi
\expandafter\ifx\csname urlprefix\endcsname\relax\def\urlprefix{URL }\fi
\providecommand{\bibinfo}[2]{#2}
\providecommand{\eprint}[2][]{\url{#2}}

\bibitem[{\citenamefont{Nagaosa and Tokura}(2013)}]{NagaosaNature}
\bibinfo{author}{\bibfnamefont{N.}~\bibnamefont{Nagaosa}} \bibnamefont{and}
  \bibinfo{author}{\bibfnamefont{Y.}~\bibnamefont{Tokura}},
  \bibinfo{journal}{Nature Nanotechnoloy} \textbf{\bibinfo{volume}{8}},
  \bibinfo{pages}{899} (\bibinfo{year}{2013}).

\bibitem[{\citenamefont{Fert et~al.}(2013)\citenamefont{Fert, Cros, and
  Sampaio}}]{fert2013skyrmions}
\bibinfo{author}{\bibfnamefont{A.}~\bibnamefont{Fert}},
  \bibinfo{author}{\bibfnamefont{V.}~\bibnamefont{Cros}}, \bibnamefont{and}
  \bibinfo{author}{\bibfnamefont{J.}~\bibnamefont{Sampaio}},
  \bibinfo{journal}{Nature nanotechnology} \textbf{\bibinfo{volume}{8}},
  \bibinfo{pages}{152} (\bibinfo{year}{2013}).

\bibitem[{\citenamefont{Fert et~al.}(2017)\citenamefont{Fert, Reyren, and
  Cros}}]{fert2017magnetic}
\bibinfo{author}{\bibfnamefont{A.}~\bibnamefont{Fert}},
  \bibinfo{author}{\bibfnamefont{N.}~\bibnamefont{Reyren}}, \bibnamefont{and}
  \bibinfo{author}{\bibfnamefont{V.}~\bibnamefont{Cros}},
  \bibinfo{journal}{Nature Reviews Materials} \textbf{\bibinfo{volume}{2}},
  \bibinfo{pages}{17031} (\bibinfo{year}{2017}).

\bibitem[{\citenamefont{Wiesendanger}(2016)}]{wiesendanger2016nanoscale}
\bibinfo{author}{\bibfnamefont{R.}~\bibnamefont{Wiesendanger}},
  \bibinfo{journal}{Nature Reviews Materials} \textbf{\bibinfo{volume}{1}},
  \bibinfo{pages}{16044} (\bibinfo{year}{2016}).

\bibitem[{\citenamefont{Ahadi et~al.}(2017)\citenamefont{Ahadi, Galletti, and
  Stemmer}}]{ahadi2017evidence}
\bibinfo{author}{\bibfnamefont{K.}~\bibnamefont{Ahadi}},
  \bibinfo{author}{\bibfnamefont{L.}~\bibnamefont{Galletti}}, \bibnamefont{and}
  \bibinfo{author}{\bibfnamefont{S.}~\bibnamefont{Stemmer}},
  \bibinfo{journal}{Applied Physics Letters} \textbf{\bibinfo{volume}{111}},
  \bibinfo{pages}{172403} (\bibinfo{year}{2017}).

\bibitem[{\citenamefont{Leroux et~al.}(2018)\citenamefont{Leroux, J.~Stolt,
  Jin, V.~Pete, Reichhardt, and Maiorov}}]{room_T}
\bibinfo{author}{\bibfnamefont{M.}~\bibnamefont{Leroux}},
  \bibinfo{author}{\bibfnamefont{M.}~\bibnamefont{J.~Stolt}},
  \bibinfo{author}{\bibfnamefont{S.}~\bibnamefont{Jin}},
  \bibinfo{author}{\bibfnamefont{D.}~\bibnamefont{V.~Pete}},
  \bibinfo{author}{\bibfnamefont{C.}~\bibnamefont{Reichhardt}},
  \bibnamefont{and} \bibinfo{author}{\bibfnamefont{B.}~\bibnamefont{Maiorov}},
  \bibinfo{journal}{arXiv:1805.05267}  (\bibinfo{year}{2018}).

\bibitem[{\citenamefont{Soumyanarayanan
  et~al.}(2017)\citenamefont{Soumyanarayanan, Raju, Oyarce, Tan, Im,
  Petrovi{\'c}, Ho, Khoo, Tran, Gan et~al.}}]{soumyanarayanan2017tunable}
\bibinfo{author}{\bibfnamefont{A.}~\bibnamefont{Soumyanarayanan}},
  \bibinfo{author}{\bibfnamefont{M.}~\bibnamefont{Raju}},
  \bibinfo{author}{\bibfnamefont{A.~G.} \bibnamefont{Oyarce}},
  \bibinfo{author}{\bibfnamefont{A.~K.} \bibnamefont{Tan}},
  \bibinfo{author}{\bibfnamefont{M.-Y.} \bibnamefont{Im}},
  \bibinfo{author}{\bibfnamefont{A.~P.} \bibnamefont{Petrovi{\'c}}},
  \bibinfo{author}{\bibfnamefont{P.}~\bibnamefont{Ho}},
  \bibinfo{author}{\bibfnamefont{K.}~\bibnamefont{Khoo}},
  \bibinfo{author}{\bibfnamefont{M.}~\bibnamefont{Tran}},
  \bibinfo{author}{\bibfnamefont{C.}~\bibnamefont{Gan}}, \bibnamefont{et~al.},
  \bibinfo{journal}{Nature materials} \textbf{\bibinfo{volume}{16}},
  \bibinfo{pages}{898} (\bibinfo{year}{2017}).

\bibitem[{\citenamefont{Raju et~al.}(2017)\citenamefont{Raju, Yagil,
  Soumyanarayanan, Tan, Almoalem, Auslaender, and
  Panagopoulos}}]{raju2017evolution}
\bibinfo{author}{\bibfnamefont{M.}~\bibnamefont{Raju}},
  \bibinfo{author}{\bibfnamefont{A.}~\bibnamefont{Yagil}},
  \bibinfo{author}{\bibfnamefont{A.}~\bibnamefont{Soumyanarayanan}},
  \bibinfo{author}{\bibfnamefont{A.~K.} \bibnamefont{Tan}},
  \bibinfo{author}{\bibfnamefont{A.}~\bibnamefont{Almoalem}},
  \bibinfo{author}{\bibfnamefont{O.}~\bibnamefont{Auslaender}},
  \bibnamefont{and}
  \bibinfo{author}{\bibfnamefont{C.}~\bibnamefont{Panagopoulos}},
  \bibinfo{journal}{arXiv preprint arXiv:1708.04084}  (\bibinfo{year}{2017}).

\bibitem[{\citenamefont{Spencer et~al.}(2018)\citenamefont{Spencer, Gayles,
  Porter, Sugimoto, Aslam, Kinane, Charlton, Freimuth, Chadov, Langridge
  et~al.}}]{B20_FeCoGe}
\bibinfo{author}{\bibfnamefont{C.~S.} \bibnamefont{Spencer}},
  \bibinfo{author}{\bibfnamefont{J.}~\bibnamefont{Gayles}},
  \bibinfo{author}{\bibfnamefont{N.~A.} \bibnamefont{Porter}},
  \bibinfo{author}{\bibfnamefont{S.}~\bibnamefont{Sugimoto}},
  \bibinfo{author}{\bibfnamefont{Z.}~\bibnamefont{Aslam}},
  \bibinfo{author}{\bibfnamefont{C.~J.} \bibnamefont{Kinane}},
  \bibinfo{author}{\bibfnamefont{T.~R.} \bibnamefont{Charlton}},
  \bibinfo{author}{\bibfnamefont{F.}~\bibnamefont{Freimuth}},
  \bibinfo{author}{\bibfnamefont{S.}~\bibnamefont{Chadov}},
  \bibinfo{author}{\bibfnamefont{S.}~\bibnamefont{Langridge}},
  \bibnamefont{et~al.}, \bibinfo{journal}{Phys. Rev. B}
  \textbf{\bibinfo{volume}{97}}, \bibinfo{pages}{214406}
  (\bibinfo{year}{2018}).

\bibitem[{\citenamefont{Neubauer et~al.}(2009)\citenamefont{Neubauer,
  Pfleiderer, Binz, Rosch, Ritz, Niklowitz, and B\"oni}}]{MnSiAPhase}
\bibinfo{author}{\bibfnamefont{A.}~\bibnamefont{Neubauer}},
  \bibinfo{author}{\bibfnamefont{C.}~\bibnamefont{Pfleiderer}},
  \bibinfo{author}{\bibfnamefont{B.}~\bibnamefont{Binz}},
  \bibinfo{author}{\bibfnamefont{A.}~\bibnamefont{Rosch}},
  \bibinfo{author}{\bibfnamefont{R.}~\bibnamefont{Ritz}},
  \bibinfo{author}{\bibfnamefont{P.~G.} \bibnamefont{Niklowitz}},
  \bibnamefont{and} \bibinfo{author}{\bibfnamefont{P.}~\bibnamefont{B\"oni}},
  \bibinfo{journal}{Phys. Rev. Lett.} \textbf{\bibinfo{volume}{102}},
  \bibinfo{pages}{186602} (\bibinfo{year}{2009}).

\bibitem[{\citenamefont{Li et~al.}(2013)\citenamefont{Li, Kanazawa, Yu,
  Tsukazaki, Kawasaki, Ichikawa, Jin, Kagawa, and Tokura}}]{THE_Li}
\bibinfo{author}{\bibfnamefont{Y.}~\bibnamefont{Li}},
  \bibinfo{author}{\bibfnamefont{N.}~\bibnamefont{Kanazawa}},
  \bibinfo{author}{\bibfnamefont{X.~Z.} \bibnamefont{Yu}},
  \bibinfo{author}{\bibfnamefont{A.}~\bibnamefont{Tsukazaki}},
  \bibinfo{author}{\bibfnamefont{M.}~\bibnamefont{Kawasaki}},
  \bibinfo{author}{\bibfnamefont{M.}~\bibnamefont{Ichikawa}},
  \bibinfo{author}{\bibfnamefont{X.~F.} \bibnamefont{Jin}},
  \bibinfo{author}{\bibfnamefont{F.}~\bibnamefont{Kagawa}}, \bibnamefont{and}
  \bibinfo{author}{\bibfnamefont{Y.}~\bibnamefont{Tokura}},
  \bibinfo{journal}{Phys. Rev. Lett.} \textbf{\bibinfo{volume}{110}},
  \bibinfo{pages}{117202} (\bibinfo{year}{2013}).

\bibitem[{\citenamefont{Chapman et~al.}(2013)\citenamefont{Chapman,
  Grossnickle, Wolf, and Lee}}]{Chapman_PRB}
\bibinfo{author}{\bibfnamefont{B.~J.} \bibnamefont{Chapman}},
  \bibinfo{author}{\bibfnamefont{M.~G.} \bibnamefont{Grossnickle}},
  \bibinfo{author}{\bibfnamefont{T.}~\bibnamefont{Wolf}}, \bibnamefont{and}
  \bibinfo{author}{\bibfnamefont{M.}~\bibnamefont{Lee}},
  \bibinfo{journal}{Phys. Rev. B} \textbf{\bibinfo{volume}{88}},
  \bibinfo{pages}{214406} (\bibinfo{year}{2013}).

\bibitem[{\citenamefont{S{\"u}rgers et~al.}(2014)\citenamefont{S{\"u}rgers,
  Fischer, Winkel, and L{\"o}hneysen}}]{surgers2014large}
\bibinfo{author}{\bibfnamefont{C.}~\bibnamefont{S{\"u}rgers}},
  \bibinfo{author}{\bibfnamefont{G.}~\bibnamefont{Fischer}},
  \bibinfo{author}{\bibfnamefont{P.}~\bibnamefont{Winkel}}, \bibnamefont{and}
  \bibinfo{author}{\bibfnamefont{H.~v.} \bibnamefont{L{\"o}hneysen}},
  \bibinfo{journal}{Nature communications} \textbf{\bibinfo{volume}{5}}
  (\bibinfo{year}{2014}).

\bibitem[{\citenamefont{Ueland et~al.}(2012)\citenamefont{Ueland, Miclea, Kato,
  Ayala-Valenzuela, McDonald, Okazaki, Tobash, Torrez, Ronning, Movshovich
  et~al.}}]{ueland2012controllable}
\bibinfo{author}{\bibfnamefont{B.}~\bibnamefont{Ueland}},
  \bibinfo{author}{\bibfnamefont{C.}~\bibnamefont{Miclea}},
  \bibinfo{author}{\bibfnamefont{Y.}~\bibnamefont{Kato}},
  \bibinfo{author}{\bibfnamefont{O.}~\bibnamefont{Ayala-Valenzuela}},
  \bibinfo{author}{\bibfnamefont{R.}~\bibnamefont{McDonald}},
  \bibinfo{author}{\bibfnamefont{R.}~\bibnamefont{Okazaki}},
  \bibinfo{author}{\bibfnamefont{P.}~\bibnamefont{Tobash}},
  \bibinfo{author}{\bibfnamefont{M.}~\bibnamefont{Torrez}},
  \bibinfo{author}{\bibfnamefont{F.}~\bibnamefont{Ronning}},
  \bibinfo{author}{\bibfnamefont{R.}~\bibnamefont{Movshovich}},
  \bibnamefont{et~al.}, \bibinfo{journal}{Nature communications}
  \textbf{\bibinfo{volume}{3}}, \bibinfo{pages}{1067} (\bibinfo{year}{2012}).

\bibitem[{\citenamefont{Fabris et~al.}(2006)\citenamefont{Fabris, Pureur,
  Schaf, Vieira, and Campbell}}]{SpinGlass1}
\bibinfo{author}{\bibfnamefont{F.~W.} \bibnamefont{Fabris}},
  \bibinfo{author}{\bibfnamefont{P.}~\bibnamefont{Pureur}},
  \bibinfo{author}{\bibfnamefont{J.}~\bibnamefont{Schaf}},
  \bibinfo{author}{\bibfnamefont{V.~N.} \bibnamefont{Vieira}},
  \bibnamefont{and} \bibinfo{author}{\bibfnamefont{I.~A.}
  \bibnamefont{Campbell}}, \bibinfo{journal}{Phys. Rev. B}
  \textbf{\bibinfo{volume}{74}}, \bibinfo{pages}{214201}
  (\bibinfo{year}{2006}).

\bibitem[{\citenamefont{Taniguchi et~al.}(2004)\citenamefont{Taniguchi,
  Yamanaka, Sumioka, Yamazaki, Tabata, and Kawarazaki}}]{SpinGlass2}
\bibinfo{author}{\bibfnamefont{T.}~\bibnamefont{Taniguchi}},
  \bibinfo{author}{\bibfnamefont{K.}~\bibnamefont{Yamanaka}},
  \bibinfo{author}{\bibfnamefont{H.}~\bibnamefont{Sumioka}},
  \bibinfo{author}{\bibfnamefont{T.}~\bibnamefont{Yamazaki}},
  \bibinfo{author}{\bibfnamefont{Y.}~\bibnamefont{Tabata}}, \bibnamefont{and}
  \bibinfo{author}{\bibfnamefont{S.}~\bibnamefont{Kawarazaki}},
  \bibinfo{journal}{Phys. Rev. Lett.} \textbf{\bibinfo{volume}{93}},
  \bibinfo{pages}{246605} (\bibinfo{year}{2004}).

\bibitem[{\citenamefont{Ohuchi et~al.}(2015)\citenamefont{Ohuchi, Kozuka,
  Uchida, Ueno, Tsukazaki, and Kawasaki}}]{EuO}
\bibinfo{author}{\bibfnamefont{Y.}~\bibnamefont{Ohuchi}},
  \bibinfo{author}{\bibfnamefont{Y.}~\bibnamefont{Kozuka}},
  \bibinfo{author}{\bibfnamefont{M.}~\bibnamefont{Uchida}},
  \bibinfo{author}{\bibfnamefont{K.}~\bibnamefont{Ueno}},
  \bibinfo{author}{\bibfnamefont{A.}~\bibnamefont{Tsukazaki}},
  \bibnamefont{and} \bibinfo{author}{\bibfnamefont{M.}~\bibnamefont{Kawasaki}},
  \bibinfo{journal}{Phys. Rev. B} \textbf{\bibinfo{volume}{91}},
  \bibinfo{pages}{245115} (\bibinfo{year}{2015}).

\bibitem[{\citenamefont{Maccariello et~al.}(2018)\citenamefont{Maccariello,
  Legrand, Reyren, Garcia, Bouzehouane, Collin, Cros, and
  Fert}}]{maccariello2018electrical}
\bibinfo{author}{\bibfnamefont{D.}~\bibnamefont{Maccariello}},
  \bibinfo{author}{\bibfnamefont{W.}~\bibnamefont{Legrand}},
  \bibinfo{author}{\bibfnamefont{N.}~\bibnamefont{Reyren}},
  \bibinfo{author}{\bibfnamefont{K.}~\bibnamefont{Garcia}},
  \bibinfo{author}{\bibfnamefont{K.}~\bibnamefont{Bouzehouane}},
  \bibinfo{author}{\bibfnamefont{S.}~\bibnamefont{Collin}},
  \bibinfo{author}{\bibfnamefont{V.}~\bibnamefont{Cros}}, \bibnamefont{and}
  \bibinfo{author}{\bibfnamefont{A.}~\bibnamefont{Fert}},
  \bibinfo{journal}{Nature Nanotechnology} pp. \bibinfo{pages}{1748--3395}
  (\bibinfo{year}{2018}).

\bibitem[{\citenamefont{Oveshnikov et~al.}(2015)\citenamefont{Oveshnikov,
  Kulbachinskii, Davydov, Aronzon, Rozhansky, Averkiev, Kugel, and
  Tripathi}}]{AronzonRozh}
\bibinfo{author}{\bibfnamefont{L.~N.} \bibnamefont{Oveshnikov}},
  \bibinfo{author}{\bibfnamefont{V.~A.} \bibnamefont{Kulbachinskii}},
  \bibinfo{author}{\bibfnamefont{A.~B.} \bibnamefont{Davydov}},
  \bibinfo{author}{\bibfnamefont{B.~A.} \bibnamefont{Aronzon}},
  \bibinfo{author}{\bibfnamefont{I.~V.} \bibnamefont{Rozhansky}},
  \bibinfo{author}{\bibfnamefont{N.~S.} \bibnamefont{Averkiev}},
  \bibinfo{author}{\bibfnamefont{K.~I.} \bibnamefont{Kugel}}, \bibnamefont{and}
  \bibinfo{author}{\bibfnamefont{V.}~\bibnamefont{Tripathi}},
  \bibinfo{journal}{Scientific Reports} \textbf{\bibinfo{volume}{5}},
  \bibinfo{pages}{17158} (\bibinfo{year}{2015}).

\bibitem[{\citenamefont{Liu et~al.}(2017)\citenamefont{Liu, Zang, Ruan, Gong,
  He, Ma, Xue, and Wang}}]{THE_TI}
\bibinfo{author}{\bibfnamefont{C.}~\bibnamefont{Liu}},
  \bibinfo{author}{\bibfnamefont{Y.}~\bibnamefont{Zang}},
  \bibinfo{author}{\bibfnamefont{W.}~\bibnamefont{Ruan}},
  \bibinfo{author}{\bibfnamefont{Y.}~\bibnamefont{Gong}},
  \bibinfo{author}{\bibfnamefont{K.}~\bibnamefont{He}},
  \bibinfo{author}{\bibfnamefont{X.}~\bibnamefont{Ma}},
  \bibinfo{author}{\bibfnamefont{Q.-K.} \bibnamefont{Xue}}, \bibnamefont{and}
  \bibinfo{author}{\bibfnamefont{Y.}~\bibnamefont{Wang}},
  \bibinfo{journal}{Phys. Rev. Lett.} \textbf{\bibinfo{volume}{119}},
  \bibinfo{pages}{176809} (\bibinfo{year}{2017}).

\bibitem[{\citenamefont{Kanazawa et~al.}(2015)\citenamefont{Kanazawa, Kubota,
  Tsukazaki, Kozuka, Takahashi, Kawasaki, Ichikawa, Kagawa, and
  Tokura}}]{DiscretHall}
\bibinfo{author}{\bibfnamefont{N.}~\bibnamefont{Kanazawa}},
  \bibinfo{author}{\bibfnamefont{M.}~\bibnamefont{Kubota}},
  \bibinfo{author}{\bibfnamefont{A.}~\bibnamefont{Tsukazaki}},
  \bibinfo{author}{\bibfnamefont{Y.}~\bibnamefont{Kozuka}},
  \bibinfo{author}{\bibfnamefont{K.~S.} \bibnamefont{Takahashi}},
  \bibinfo{author}{\bibfnamefont{M.}~\bibnamefont{Kawasaki}},
  \bibinfo{author}{\bibfnamefont{M.}~\bibnamefont{Ichikawa}},
  \bibinfo{author}{\bibfnamefont{F.}~\bibnamefont{Kagawa}}, \bibnamefont{and}
  \bibinfo{author}{\bibfnamefont{Y.}~\bibnamefont{Tokura}},
  \bibinfo{journal}{Phys. Rev. B} \textbf{\bibinfo{volume}{91}},
  \bibinfo{pages}{041122} (\bibinfo{year}{2015}).

\bibitem[{\citenamefont{Mühlbauer et~al.}(2009)\citenamefont{Mühlbauer, Binz,
  Jonietz, Pfleiderer, Rosch, Neubauer, Georgii, and
  Böni}}]{Muhl_MnSi_Science}
\bibinfo{author}{\bibfnamefont{S.}~\bibnamefont{Mühlbauer}},
  \bibinfo{author}{\bibfnamefont{B.}~\bibnamefont{Binz}},
  \bibinfo{author}{\bibfnamefont{F.}~\bibnamefont{Jonietz}},
  \bibinfo{author}{\bibfnamefont{C.}~\bibnamefont{Pfleiderer}},
  \bibinfo{author}{\bibfnamefont{A.}~\bibnamefont{Rosch}},
  \bibinfo{author}{\bibfnamefont{A.}~\bibnamefont{Neubauer}},
  \bibinfo{author}{\bibfnamefont{R.}~\bibnamefont{Georgii}}, \bibnamefont{and}
  \bibinfo{author}{\bibfnamefont{P.}~\bibnamefont{Böni}},
  \bibinfo{journal}{Science} \textbf{\bibinfo{volume}{323}},
  \bibinfo{pages}{915} (\bibinfo{year}{2009}).

\bibitem[{\citenamefont{M\"unzer et~al.}(2010)\citenamefont{M\"unzer, Neubauer,
  Adams, M\"uhlbauer, Franz, Jonietz, Georgii, B\"oni, Pedersen, Schmidt
  et~al.}}]{Munzer_PRB}
\bibinfo{author}{\bibfnamefont{W.}~\bibnamefont{M\"unzer}},
  \bibinfo{author}{\bibfnamefont{A.}~\bibnamefont{Neubauer}},
  \bibinfo{author}{\bibfnamefont{T.}~\bibnamefont{Adams}},
  \bibinfo{author}{\bibfnamefont{S.}~\bibnamefont{M\"uhlbauer}},
  \bibinfo{author}{\bibfnamefont{C.}~\bibnamefont{Franz}},
  \bibinfo{author}{\bibfnamefont{F.}~\bibnamefont{Jonietz}},
  \bibinfo{author}{\bibfnamefont{R.}~\bibnamefont{Georgii}},
  \bibinfo{author}{\bibfnamefont{P.}~\bibnamefont{B\"oni}},
  \bibinfo{author}{\bibfnamefont{B.}~\bibnamefont{Pedersen}},
  \bibinfo{author}{\bibfnamefont{M.}~\bibnamefont{Schmidt}},
  \bibnamefont{et~al.}, \bibinfo{journal}{Phys. Rev. B}
  \textbf{\bibinfo{volume}{81}}, \bibinfo{pages}{041203}
  (\bibinfo{year}{2010}).

\bibitem[{\citenamefont{Bruno et~al.}(2004)\citenamefont{Bruno, Dugaev, and
  Taillefumier}}]{BrunoDugaev}
\bibinfo{author}{\bibfnamefont{P.}~\bibnamefont{Bruno}},
  \bibinfo{author}{\bibfnamefont{V.~K.} \bibnamefont{Dugaev}},
  \bibnamefont{and}
  \bibinfo{author}{\bibfnamefont{M.}~\bibnamefont{Taillefumier}},
  \bibinfo{journal}{Phys. Rev. Lett.} \textbf{\bibinfo{volume}{93}},
  \bibinfo{pages}{096806} (\bibinfo{year}{2004}).

\bibitem[{\citenamefont{Binz and Vishwanath}(2008)}]{binz2008chirality}
\bibinfo{author}{\bibfnamefont{B.}~\bibnamefont{Binz}} \bibnamefont{and}
  \bibinfo{author}{\bibfnamefont{A.}~\bibnamefont{Vishwanath}},
  \bibinfo{journal}{Physica B: Condensed Matter}
  \textbf{\bibinfo{volume}{403}}, \bibinfo{pages}{1336} (\bibinfo{year}{2008}).

\bibitem[{\citenamefont{Ye et~al.}(1999)\citenamefont{Ye, Kim, Millis,
  Shraiman, Majumdar, and Te\ifmmode \check{s}\else
  \v{s}\fi{}anovi\ifmmode~\acute{c}\else \'{c}\fi{}}}]{Ye1999}
\bibinfo{author}{\bibfnamefont{J.}~\bibnamefont{Ye}},
  \bibinfo{author}{\bibfnamefont{Y.~B.} \bibnamefont{Kim}},
  \bibinfo{author}{\bibfnamefont{A.~J.} \bibnamefont{Millis}},
  \bibinfo{author}{\bibfnamefont{B.~I.} \bibnamefont{Shraiman}},
  \bibinfo{author}{\bibfnamefont{P.}~\bibnamefont{Majumdar}}, \bibnamefont{and}
  \bibinfo{author}{\bibfnamefont{Z.}~\bibnamefont{Te\ifmmode \check{s}\else
  \v{s}\fi{}anovi\ifmmode~\acute{c}\else \'{c}\fi{}}}, \bibinfo{journal}{Phys.
  Rev. Lett.} \textbf{\bibinfo{volume}{83}}, \bibinfo{pages}{3737}
  (\bibinfo{year}{1999}).

\bibitem[{\citenamefont{Lyanda-Geller et~al.}(2001)\citenamefont{Lyanda-Geller,
  Chun, Salamon, Goldbart, Han, Tomioka, Asamitsu, and Tokura}}]{Lyana-Geller1}
\bibinfo{author}{\bibfnamefont{Y.}~\bibnamefont{Lyanda-Geller}},
  \bibinfo{author}{\bibfnamefont{S.~H.} \bibnamefont{Chun}},
  \bibinfo{author}{\bibfnamefont{M.~B.} \bibnamefont{Salamon}},
  \bibinfo{author}{\bibfnamefont{P.~M.} \bibnamefont{Goldbart}},
  \bibinfo{author}{\bibfnamefont{P.~D.} \bibnamefont{Han}},
  \bibinfo{author}{\bibfnamefont{Y.}~\bibnamefont{Tomioka}},
  \bibinfo{author}{\bibfnamefont{A.}~\bibnamefont{Asamitsu}}, \bibnamefont{and}
  \bibinfo{author}{\bibfnamefont{Y.}~\bibnamefont{Tokura}},
  \bibinfo{journal}{Phys. Rev. B} \textbf{\bibinfo{volume}{63}},
  \bibinfo{pages}{184426} (\bibinfo{year}{2001}).

\bibitem[{\citenamefont{Ndiaye et~al.}(2017)\citenamefont{Ndiaye, Akosa, and
  Manchon}}]{Arab_Papa}
\bibinfo{author}{\bibfnamefont{P.~B.} \bibnamefont{Ndiaye}},
  \bibinfo{author}{\bibfnamefont{C.~A.} \bibnamefont{Akosa}}, \bibnamefont{and}
  \bibinfo{author}{\bibfnamefont{A.}~\bibnamefont{Manchon}},
  \bibinfo{journal}{Phys. Rev. B} \textbf{\bibinfo{volume}{95}},
  \bibinfo{pages}{064426} (\bibinfo{year}{2017}).

\bibitem[{\citenamefont{Tatara et~al.}(2007)\citenamefont{Tatara, Kohno,
  Shibata, Lemaho, and Lee}}]{Tatara_2007}
\bibinfo{author}{\bibfnamefont{G.}~\bibnamefont{Tatara}},
  \bibinfo{author}{\bibfnamefont{H.}~\bibnamefont{Kohno}},
  \bibinfo{author}{\bibfnamefont{J.}~\bibnamefont{Shibata}},
  \bibinfo{author}{\bibfnamefont{Y.}~\bibnamefont{Lemaho}}, \bibnamefont{and}
  \bibinfo{author}{\bibfnamefont{K.-J.} \bibnamefont{Lee}},
  \bibinfo{journal}{Journal of the Physical Society of Japan}
  \textbf{\bibinfo{volume}{76}}, \bibinfo{pages}{054707}
  (\bibinfo{year}{2007}).

\bibitem[{\citenamefont{Zhang and Heinonen}(2018)}]{Diffusive_THE}
\bibinfo{author}{\bibfnamefont{S.~S.-L.} \bibnamefont{Zhang}} \bibnamefont{and}
  \bibinfo{author}{\bibfnamefont{O.}~\bibnamefont{Heinonen}},
  \bibinfo{journal}{Phys. Rev. B} \textbf{\bibinfo{volume}{97}},
  \bibinfo{pages}{134401} (\bibinfo{year}{2018}).

\bibitem[{\citenamefont{Taguchi et~al.}(2001)\citenamefont{Taguchi, Oohara,
  Yoshizawa, Nagaosa, and Tokura}}]{Taguchi_Science}
\bibinfo{author}{\bibfnamefont{Y.}~\bibnamefont{Taguchi}},
  \bibinfo{author}{\bibfnamefont{Y.}~\bibnamefont{Oohara}},
  \bibinfo{author}{\bibfnamefont{H.}~\bibnamefont{Yoshizawa}},
  \bibinfo{author}{\bibfnamefont{N.}~\bibnamefont{Nagaosa}}, \bibnamefont{and}
  \bibinfo{author}{\bibfnamefont{Y.}~\bibnamefont{Tokura}},
  \bibinfo{journal}{Science} \textbf{\bibinfo{volume}{291}},
  \bibinfo{pages}{2573} (\bibinfo{year}{2001}).

\bibitem[{\citenamefont{Nakazawa et~al.}(2018)\citenamefont{Nakazawa, Bibes,
  and Kohno}}]{nakazawa2018topological}
\bibinfo{author}{\bibfnamefont{K.}~\bibnamefont{Nakazawa}},
  \bibinfo{author}{\bibfnamefont{M.}~\bibnamefont{Bibes}}, \bibnamefont{and}
  \bibinfo{author}{\bibfnamefont{H.}~\bibnamefont{Kohno}},
  \bibinfo{journal}{Journal of the Physical Society of Japan}
  \textbf{\bibinfo{volume}{87}}, \bibinfo{pages}{033705}
  (\bibinfo{year}{2018}).

\bibitem[{\citenamefont{Ishizuka and Nagaosa}(2018)}]{ishizuka2018spin}
\bibinfo{author}{\bibfnamefont{H.}~\bibnamefont{Ishizuka}} \bibnamefont{and}
  \bibinfo{author}{\bibfnamefont{N.}~\bibnamefont{Nagaosa}},
  \bibinfo{journal}{Science Advances} \textbf{\bibinfo{volume}{4}},
  \bibinfo{pages}{eaap9962} (\bibinfo{year}{2018}).

\bibitem[{\citenamefont{Denisov and Averkiev}(2018)}]{denisov2018hall}
\bibinfo{author}{\bibfnamefont{K.}~\bibnamefont{Denisov}} \bibnamefont{and}
  \bibinfo{author}{\bibfnamefont{N.}~\bibnamefont{Averkiev}},
  \bibinfo{journal}{Applied Physics Letters} \textbf{\bibinfo{volume}{112}},
  \bibinfo{pages}{162409} (\bibinfo{year}{2018}).

\bibitem[{\citenamefont{Denisov et~al.}(2017)\citenamefont{Denisov, Rozhansky,
  Averkiev, and L\"ahderanta}}]{SciRep_skyrmion}
\bibinfo{author}{\bibfnamefont{K.~S.} \bibnamefont{Denisov}},
  \bibinfo{author}{\bibfnamefont{I.~V.} \bibnamefont{Rozhansky}},
  \bibinfo{author}{\bibfnamefont{N.~S.} \bibnamefont{Averkiev}},
  \bibnamefont{and}
  \bibinfo{author}{\bibfnamefont{E.}~\bibnamefont{L\"ahderanta}},
  \bibinfo{journal}{Scientific Reports} \textbf{\bibinfo{volume}{7}},
  \bibinfo{pages}{17204} (\bibinfo{year}{2017}).

\bibitem[{\citenamefont{Denisov et~al.}(2016)\citenamefont{Denisov, Rozhansky,
  Averkiev, and L\"ahderanta}}]{prl_skyrmion}
\bibinfo{author}{\bibfnamefont{K.~S.} \bibnamefont{Denisov}},
  \bibinfo{author}{\bibfnamefont{I.~V.} \bibnamefont{Rozhansky}},
  \bibinfo{author}{\bibfnamefont{N.~S.} \bibnamefont{Averkiev}},
  \bibnamefont{and}
  \bibinfo{author}{\bibfnamefont{E.}~\bibnamefont{L\"ahderanta}},
  \bibinfo{journal}{Phys. Rev. Lett.} \textbf{\bibinfo{volume}{117}},
  \bibinfo{pages}{027202} (\bibinfo{year}{2016}).

\bibitem[{\citenamefont{Tatara and Kawamura}(2002)}]{Tatara}
\bibinfo{author}{\bibfnamefont{G.}~\bibnamefont{Tatara}} \bibnamefont{and}
  \bibinfo{author}{\bibfnamefont{H.}~\bibnamefont{Kawamura}},
  \bibinfo{journal}{J. Phys. Soc. of Japan} \textbf{\bibinfo{volume}{71}},
  \bibinfo{pages}{2613} (\bibinfo{year}{2002}).

\bibitem[{\citenamefont{Nakazawa and Kohno}(2014)}]{Nakazawa2014}
\bibinfo{author}{\bibfnamefont{K.}~\bibnamefont{Nakazawa}} \bibnamefont{and}
  \bibinfo{author}{\bibfnamefont{H.}~\bibnamefont{Kohno}}, \bibinfo{journal}{J.
  Phys. Soc. of Japan} \textbf{\bibinfo{volume}{83}}, \bibinfo{pages}{073707}
  (\bibinfo{year}{2014}).

\bibitem[{\citenamefont{Onoda et~al.}(2004)\citenamefont{Onoda, Tatara, and
  Nagaosa}}]{Onoda_SkyrmNumber}
\bibinfo{author}{\bibfnamefont{M.}~\bibnamefont{Onoda}},
  \bibinfo{author}{\bibfnamefont{G.}~\bibnamefont{Tatara}}, \bibnamefont{and}
  \bibinfo{author}{\bibfnamefont{N.}~\bibnamefont{Nagaosa}},
  \bibinfo{journal}{Journal of the Physical Society of Japan}
  \textbf{\bibinfo{volume}{73}}, \bibinfo{pages}{2624} (\bibinfo{year}{2004}).

\bibitem[{\citenamefont{Lobanov et~al.}(2016)\citenamefont{Lobanov,
  J{\'o}nsson, and Uzdin}}]{Uzdin-1}
\bibinfo{author}{\bibfnamefont{I.~S.} \bibnamefont{Lobanov}},
  \bibinfo{author}{\bibfnamefont{H.}~\bibnamefont{J{\'o}nsson}},
  \bibnamefont{and} \bibinfo{author}{\bibfnamefont{V.~M.} \bibnamefont{Uzdin}},
  \bibinfo{journal}{Physical Review B} \textbf{\bibinfo{volume}{94}},
  \bibinfo{pages}{174418} (\bibinfo{year}{2016}).

\bibitem[{\citenamefont{Moreau-Luchaire
  et~al.}(2016)\citenamefont{Moreau-Luchaire, Moutafis, Reyren, Sampaio, Vaz,
  Van~Horne, Bouzehouane, Garcia, Deranlot, Warnicke
  et~al.}}]{moreau2016additive}
\bibinfo{author}{\bibfnamefont{C.}~\bibnamefont{Moreau-Luchaire}},
  \bibinfo{author}{\bibfnamefont{C.}~\bibnamefont{Moutafis}},
  \bibinfo{author}{\bibfnamefont{N.}~\bibnamefont{Reyren}},
  \bibinfo{author}{\bibfnamefont{J.}~\bibnamefont{Sampaio}},
  \bibinfo{author}{\bibfnamefont{C.}~\bibnamefont{Vaz}},
  \bibinfo{author}{\bibfnamefont{N.}~\bibnamefont{Van~Horne}},
  \bibinfo{author}{\bibfnamefont{K.}~\bibnamefont{Bouzehouane}},
  \bibinfo{author}{\bibfnamefont{K.}~\bibnamefont{Garcia}},
  \bibinfo{author}{\bibfnamefont{C.}~\bibnamefont{Deranlot}},
  \bibinfo{author}{\bibfnamefont{P.}~\bibnamefont{Warnicke}},
  \bibnamefont{et~al.}, \bibinfo{journal}{Nature nanotechnology}
  \textbf{\bibinfo{volume}{11}}, \bibinfo{pages}{444} (\bibinfo{year}{2016}).

\bibitem[{\citenamefont{Soumyanarayanan
  et~al.}(2016)\citenamefont{Soumyanarayanan, Reyren, Fert, and
  Panagopoulos}}]{soumyanarayanan2016emergent}
\bibinfo{author}{\bibfnamefont{A.}~\bibnamefont{Soumyanarayanan}},
  \bibinfo{author}{\bibfnamefont{N.}~\bibnamefont{Reyren}},
  \bibinfo{author}{\bibfnamefont{A.}~\bibnamefont{Fert}}, \bibnamefont{and}
  \bibinfo{author}{\bibfnamefont{C.}~\bibnamefont{Panagopoulos}},
  \bibinfo{journal}{Nature} \textbf{\bibinfo{volume}{539}},
  \bibinfo{pages}{509} (\bibinfo{year}{2016}).

\bibitem[{\citenamefont{Bogdanov and
  Hubert}(1994)}]{bogdanov1994thermodynamically}
\bibinfo{author}{\bibfnamefont{A.}~\bibnamefont{Bogdanov}} \bibnamefont{and}
  \bibinfo{author}{\bibfnamefont{A.}~\bibnamefont{Hubert}},
  \bibinfo{journal}{Journal of magnetism and magnetic materials}
  \textbf{\bibinfo{volume}{138}}, \bibinfo{pages}{255} (\bibinfo{year}{1994}).

\bibitem[{\citenamefont{Bogdanov and Hubert}(1999)}]{bogdanov1999stability}
\bibinfo{author}{\bibfnamefont{A.}~\bibnamefont{Bogdanov}} \bibnamefont{and}
  \bibinfo{author}{\bibfnamefont{A.}~\bibnamefont{Hubert}},
  \bibinfo{journal}{Journal of magnetism and magnetic materials}
  \textbf{\bibinfo{volume}{195}}, \bibinfo{pages}{182} (\bibinfo{year}{1999}).

\bibitem[{\citenamefont{Belavin and Polyakov}(1975)}]{belavin1975metastable}
\bibinfo{author}{\bibfnamefont{A.}~\bibnamefont{Belavin}} \bibnamefont{and}
  \bibinfo{author}{\bibfnamefont{A.}~\bibnamefont{Polyakov}},
  \bibinfo{journal}{JETP lett} \textbf{\bibinfo{volume}{22}},
  \bibinfo{pages}{245} (\bibinfo{year}{1975}).

\bibitem[{\citenamefont{Brey}(2017)}]{brey2017magnetic}
\bibinfo{author}{\bibfnamefont{L.}~\bibnamefont{Brey}}, \bibinfo{journal}{Nano
  letters} \textbf{\bibinfo{volume}{17}}, \bibinfo{pages}{7358}
  (\bibinfo{year}{2017}).

\bibitem[{\citenamefont{Bessarab et~al.}(2018)\citenamefont{Bessarab,
  M{\"u}ller, Lobanov, Rybakov, Kiselev, J{\'o}nsson, Uzdin, Bl{\"u}gel,
  Bergqvist, and Delin}}]{bessarab2018lifetime}
\bibinfo{author}{\bibfnamefont{P.~F.} \bibnamefont{Bessarab}},
  \bibinfo{author}{\bibfnamefont{G.~P.} \bibnamefont{M{\"u}ller}},
  \bibinfo{author}{\bibfnamefont{I.~S.} \bibnamefont{Lobanov}},
  \bibinfo{author}{\bibfnamefont{F.~N.} \bibnamefont{Rybakov}},
  \bibinfo{author}{\bibfnamefont{N.~S.} \bibnamefont{Kiselev}},
  \bibinfo{author}{\bibfnamefont{H.}~\bibnamefont{J{\'o}nsson}},
  \bibinfo{author}{\bibfnamefont{V.~M.} \bibnamefont{Uzdin}},
  \bibinfo{author}{\bibfnamefont{S.}~\bibnamefont{Bl{\"u}gel}},
  \bibinfo{author}{\bibfnamefont{L.}~\bibnamefont{Bergqvist}},
  \bibnamefont{and} \bibinfo{author}{\bibfnamefont{A.}~\bibnamefont{Delin}},
  \bibinfo{journal}{Scientific reports} \textbf{\bibinfo{volume}{8}},
  \bibinfo{pages}{3433} (\bibinfo{year}{2018}).

\bibitem[{\citenamefont{Denisov and Averkiev}(2014)}]{DenisovPolaron}
\bibinfo{author}{\bibfnamefont{K.~S.} \bibnamefont{Denisov}} \bibnamefont{and}
  \bibinfo{author}{\bibfnamefont{N.~S.} \bibnamefont{Averkiev}},
  \bibinfo{journal}{JETP Letters} \textbf{\bibinfo{volume}{99}},
  \bibinfo{pages}{400} (\bibinfo{year}{2014}).

\bibitem[{\citenamefont{Berkovskaya et~al.}(1998)\citenamefont{Berkovskaya,
  Gel'mont, and Tsidil'kovskii}}]{Berk}
\bibinfo{author}{\bibfnamefont{Y.}~\bibnamefont{Berkovskaya}},
  \bibinfo{author}{\bibfnamefont{B.}~\bibnamefont{Gel'mont}}, \bibnamefont{and}
  \bibinfo{author}{\bibfnamefont{E.}~\bibnamefont{Tsidil'kovskii}},
  \bibinfo{journal}{Sov. Phys.-Semiconductors} \textbf{\bibinfo{volume}{22}},
  \bibinfo{pages}{539} (\bibinfo{year}{1998}).

\bibitem[{\citenamefont{Berkovskaya et~al.}(1988)\citenamefont{Berkovskaya,
  Vakhabova, Gel'mont, and Merkulov}}]{Merkulov_Polaron}
\bibinfo{author}{\bibfnamefont{Y.}~\bibnamefont{Berkovskaya}},
  \bibinfo{author}{\bibfnamefont{E.}~\bibnamefont{Vakhabova}},
  \bibinfo{author}{\bibfnamefont{B.}~\bibnamefont{Gel'mont}}, \bibnamefont{and}
  \bibinfo{author}{\bibfnamefont{I.}~\bibnamefont{Merkulov}},
  \bibinfo{journal}{JETP} \textbf{\bibinfo{volume}{67}}, \bibinfo{pages}{750}
  (\bibinfo{year}{1988}).

\bibitem[{\citenamefont{Denisov et~al.}(2018)\citenamefont{Denisov, Rozhansky,
  Potkina, Lobanov, Lahderanta, and Uzdin}}]{denisov2018topological}
\bibinfo{author}{\bibfnamefont{K.}~\bibnamefont{Denisov}},
  \bibinfo{author}{\bibfnamefont{I.}~\bibnamefont{Rozhansky}},
  \bibinfo{author}{\bibfnamefont{M.}~\bibnamefont{Potkina}},
  \bibinfo{author}{\bibfnamefont{I.}~\bibnamefont{Lobanov}},
  \bibinfo{author}{\bibfnamefont{E.}~\bibnamefont{Lahderanta}},
  \bibnamefont{and} \bibinfo{author}{\bibfnamefont{V.}~\bibnamefont{Uzdin}},
  \bibinfo{journal}{arXiv preprint arXiv:1803.06124}  (\bibinfo{year}{2018}).

\bibitem[{\citenamefont{Moca et~al.}(2009)\citenamefont{Moca, Sheu, Samarth,
  Schiffer, Janko, and Zarand}}]{Moca2009}
\bibinfo{author}{\bibfnamefont{C.~P.} \bibnamefont{Moca}},
  \bibinfo{author}{\bibfnamefont{B.~L.} \bibnamefont{Sheu}},
  \bibinfo{author}{\bibfnamefont{N.}~\bibnamefont{Samarth}},
  \bibinfo{author}{\bibfnamefont{P.}~\bibnamefont{Schiffer}},
  \bibinfo{author}{\bibfnamefont{B.}~\bibnamefont{Janko}}, \bibnamefont{and}
  \bibinfo{author}{\bibfnamefont{G.}~\bibnamefont{Zarand}},
  \bibinfo{journal}{Phys. Rev. Lett.} \textbf{\bibinfo{volume}{102}},
  \bibinfo{pages}{137203} (\bibinfo{year}{2009}).

\bibitem[{\citenamefont{Kamal and Murthy}(1993)}]{kamal1993new}
\bibinfo{author}{\bibfnamefont{M.}~\bibnamefont{Kamal}} \bibnamefont{and}
  \bibinfo{author}{\bibfnamefont{G.}~\bibnamefont{Murthy}},
  \bibinfo{journal}{Physical review letters} \textbf{\bibinfo{volume}{71}},
  \bibinfo{pages}{1911} (\bibinfo{year}{1993}).

\end{thebibliography}


\end{document}